\documentclass[10pt]{article}

\newlength{\minitwocolumn}
\font\teneufm=eufm10
\font\seveneufm=eufm7
\font\fiveeufm=eufm5
\newfam\eufmfam
\textfont\eufmfam=\teneufm
\scriptfont\eufmfam=\seveneufm
\scriptscriptfont\eufmfam=\fiveeufm

\makeatletter
\@addtoreset{equation}{section}
\makeatother

\newtheorem{thm}{Theorem}[section]
\newtheorem{prop}[thm]{Proposition}

\newtheorem{dfn}[thm]{Definition}
\title{\bf
\Large{\bf Boundary state of\\
$U_q(\widehat{gl}(N|N))$ analog of half-infinite $t-J$ model}}
\author{}
\begin{document}

\maketitle

\begin{center}
{Takeo Kojima}
\\~\\
{\it
Department of Mathematics and Physics,
Faculty of Engineering,
Yamagata University,\\
Jonan 4-3-16, Yonezawa 992-8510, JAPAN}
\\
kojima@yz.yamagata-u.ac.jp
\end{center}

~\\
~\\

\begin{abstract}
The $U_q(\widehat{gl}(N|N))$-analog 
of the half-infinite $t-J$ model with a boundary
is considered by using the vertex operator approach.
We find explicit bosonic formula of
the boundary state in the integrable highest-weight module
over the quantum superalgebra $U_q(\widehat{gl}(N|N))$.
\end{abstract}

~\\
~\\

\section{Introduction}
There have been many developments in the exactly solvable models.
Various methods were invented to solve models.
The vertex operator approach \cite{JM, JKKKM} provides a powerful method to study
exactly solvable models in the thermodynamic limit.
This paper is devoted to the vertex operator approach to half-infinite lattice
with open boundary.
Exactly solvable lattice models with open boundary are defined by
using the Yang-Baxter equation
and the boundary Yang-Baxter equation \cite{C, S}
\begin{eqnarray}
K_2(z_2)R_{2,1}(z_1z_2)K_1(z_1)R_{1,2}(z_1/z_2)=
R_{2,1}(z_1/z_2)K_1(z_1)R_{1,2}(z_1z_2)K_2(z_2).\nonumber
\end{eqnarray}
The vertex operator approach to the half-infinite
lattice have been studied for the quantum affine algebras 
$U_q(\widehat{sl}(N))$, $U_q(A_2^{(2)})$, $U_q(\widehat{sl}(M|N))~(M \neq N)$ \cite{JKKKM, FK, YZ1, K1, K3}
and
the elliptic deformed algebra $U_{q,p}(\widehat{sl}(N))$ \cite{MW, K2}.
The author \cite{K3} considered 
the $U_q(\widehat{sl}(M|N))$-analog of the half-infinite $t-J$ model with a boundary, and 
found explicit bosonic formula for the boundary state
in the irreducible highest-weight module.
However the very interesting case of $M=N$ has been ignored.
The quantum superalgebra $U_q(\widehat{gl}(N|N))$ is
the only untwisted superalgebra which has nonstandard system where all simple roots are odd or fermionic.
In this paper we study
the $U_q(\widehat{gl}(N|N))$-analog 
of the half-infinite $t-J$ model with a boundary, 
using the vertex operator approach.
The boundary condition of our model
is given by general diagonal solution of the boundary Yang-Baxter equation
\begin{eqnarray}
\overline{K}(z)={\rm diag}\left(z^2,\cdots,z^2,\frac{1-rz}{1-r/z},\cdots,\frac{1-rz}{1-r/z},1,\cdots,1\right).\nonumber
\end{eqnarray}
In the vertex operator approach, 
transfer matrix $T_B(z)$ of solvable lattice model with a boundary
is written by vertex operators $\Phi_j(z)$, $\Phi_j^*(z)$, and a solution of the boundary Yang-Baxter equation
as follows.
\begin{eqnarray}
T_B(z)=g \sum_{j,k=1}^{2N}
\Phi_j^*(z^{-1})K(z)_j^k \Phi_k(z) (-1)^{[v_j]}.\nonumber
\end{eqnarray}
We are interested in a realization of the eigenvector ${_B\langle i|}$ that satisfies
\begin{eqnarray}
{_B\langle i|}T_B(z)={_B \langle i|}.\nonumber
\end{eqnarray}
We call this eigenvector ${_B\langle i|}$ the boundary state.
The boundary state ${_B\langle i|}$ 
is realized by acting exponential of the bosonic operator $G$
on the highest-weight vector $\langle \Lambda_i|$
in the integrable highest-weight module $V^*(\Lambda_i)$.
\begin{eqnarray}
{_B\langle i|}=\langle \Lambda_i |e^{G} \cdot {Pr}.\nonumber
\end{eqnarray} 
Here $Pr$ is the projection operator.

The text is organized as follows.
In Section \ref{sec:2} we introduce the $R$-matrix and the boundary $K$-matrix.
We introduce
the $U_q(\widehat{gl}(N|N))$-analog 
of the half-infinite $t-J$ model with a boundary.
In Section \ref{sec:3},
we formulate the vertex operator approach to our problem,
which is free from difficulty of divergence. 
In Section \ref{sec:4} we review a bosonization of 
the quantum superalgebra $U_q(\widehat{gl}(N|N))$ and integral representation of the vertex operator.
In Section \ref{sec:5} we give a bosonization of the boundary state in 
the integrable highest-weight module $V^*(\Lambda_0)$.
In Section \ref{sec:6}
we give a proof of characterizing relation of the boundary state
by using integral representation of the vertex operator.
In Section \ref{sec:7}
we discuss generalizations of the present paper.
In Appendix \ref{app:1}
we review the quantum superalgebra $U_q(\widehat{gl}(N|N))$.
In Appendix \ref{app:2}
we give a bosonization of the boundary state in the integrable highest-weight module $V^*(\Lambda_{2N-1})$.
In Appendix \ref{app:3}
we summarize normal orderings of fundamental bosonic fields.

\section{$U_q(\widehat{gl}(N|N))$-analog of half-infinite $t-J$ model}
\label{sec:2}
In this Section we introduce
the $U_q(\widehat{gl}(N|N))$-analog 
of the half-infinite $t-J$ model with a boundary.

\subsection{$R$-matrix and $K$-matrix}
In this Section we introduce the $R$-matrix and the boundary $K$-matrix.
Let $N \in {\bf N}_{\neq 0}$ and $q \in {\bf C}$ such that $0<|q|<1$.
Let $L,M,R \in {\bf N}$ such that $L+M+R=2N$.
We set the vector space $V=\oplus_{j=1}^{2N} {\bf C}v_j$.
The ${\bf Z}_2$-grading of the basis $\{v_j\}_{1\leq j \leq 2N}$ is given to be $[v_j]=\left\{\begin{array}{cc}
0 & (j={\rm odd})\\
1 & (j={\rm even})
\end{array}\right.$.
The ${\bf Z}_2$-grading of matrix $A=(A_{j,k})_{1\leq j,k \leq 2N} \in {\rm End}(V)$ 
is defined by $[A]=[v_j]+[v_k]~({mod.} 2)$ if RHS of the equation does not depend on $j$ and $k$
such that $A_{j,k}\neq 0$.
We define action of operator $A_1\otimes A_2 \otimes \cdots \otimes A_n$ where $A_j\in {\rm End}(V)$ have ${\bf Z}_2$-grading.
We set
\begin{eqnarray}
&&
A_1 \otimes A_2 \otimes \cdots \otimes A_n \cdot v_{j_1}\otimes v_{j_2}\otimes \cdots \otimes v_{j_n}\nonumber 
\\
&=&\exp\left(\pi \sqrt{-1}\sum_{k=1}^n [A_k]\sum_{s=1}^{k-1}[v_{j_s}]\right)
A_1 v_{j_1}\otimes A_2 v_{j_2} \otimes \cdots \otimes A_n v_{j_n}.
\end{eqnarray}
We have the following multiplication rule.
\begin{eqnarray}
(A_1 \otimes A_2) (B_1\otimes B_2)=(-1)^{[A_2][B_1]}(A_1 B_1\otimes A_2 B_2).
\end{eqnarray}
We introduce super-trace "str" and super-transpose "st" of $A \in {\rm End}(V)$ by
\begin{eqnarray}
{\rm str}_V(A)=\sum_{j=1}^{2N} (-1)^{[v_j]}A_{j,j},~~~
(A^{st})_{j,k}=A_{k,j}(-1)^{[v_j]([v_j]+[v_k])}.
\end{eqnarray}

\begin{dfn}~~~Let ${R}(z) \in {\rm End}(V \otimes V)$ be the $R$-matrix of
$U_q(\widehat{gl}(N|N))$ defined by
\begin{eqnarray}
R(z)=r(z)\overline{R}(z),~~~\overline{R}(z) v_i\otimes v_j=\sum_{k,l=1}^{2N}
v_k \otimes v_l \overline{R}_{k,l}^{i,j}(z),
\label{def:R-matrix}
\end{eqnarray}
where
\begin{eqnarray}
&&
\overline{R}_{2j-1,2j-1}^{2j-1,2j-1}(z)=1,~~~\overline{R}_{2j,2j}^{2j,2j}(z)=\frac{z-q^2}{q^2z-1}~~~(1\leq j \leq N),\\
&&
\overline{R}_{i,j}^{i,j}(z)=\frac{q(z-1)}{q^2z-1}~~~(1\leq i \neq j \leq 2N),\\
&&
\overline{R}_{i,j}^{j,i}(z)=\frac{q^2-1}{q^2z-1}(-1)^{[v_i][v_j]},~~~
\overline{R}_{j,i}^{i,j}(z)=\frac{(q^2-1)z}{q^2z-1}(-1)^{[v_i][v_j]}~~~(1\leq i<j \leq 2N),\\
&&
\overline{R}_{k,l}^{i,j}(z)=0~~~({\rm otherwise}).
\end{eqnarray}
The scalar function $r(z)$ in (\ref{def:R-matrix}) is
\begin{eqnarray}
r(z)=-z^{\frac{1-2N}{N}}\left(\frac{1-q/z}{1-qz}\right)^{\frac{N-1}{N}}.
\end{eqnarray}
\end{dfn}
The $R$-matrix $R(z)$ satisfies the graded Yang-Baxter equation in $V \otimes V \otimes V$.
\begin{eqnarray}
R_{1,2}(z_1/z_2)R_{1,3}(z_1/z_3)R_{2,3}(z_2/z_3)=
R_{2,3}(z_2/z_3)R_{1,3}(z_1/z_3)R_{1,2}(z_1/z_2).
\end{eqnarray}
The $R$-matrix $R(z)$ satisfies (i) initial condition $R(1)=P$ with $P$ being the graded permutation operator 
$P_{i,j}^{k,l}=\delta_{i,k}\delta_{j,l}(-1)^{[v_i][v_j]}$, 
(ii) unitary condition $R_{1,2}(z)R_{2,1}(z^{-1})=1$, and (iii) crossing symmetry
$(R(z)^{-1})^{st_1} R(z)^{st_1}=\frac{(1-z)^2}{(1-z/q^2)(1-q^2z)}$.

\begin{dfn}~~~
Let ${K}(z) \in {\rm End}(V)$ be the boundary $K$-matrix of
$U_q(\widehat{gl}(N|N))$ defined by
\begin{eqnarray}
K(z)=\frac{\varphi(z)}{\varphi(z^{-1})}
\overline{K}(z),~~~
\overline{K}(z) v_i=\sum_{j=1}^{2N}
v_j \overline{K}_{j}^{i}(z),
\label{def:K-matrix}
\end{eqnarray}
where diagonal matrix $\overline{K}(z)$ is defined by
\begin{eqnarray}
\overline{K}(z)=
{\rm diag}\left(\underbrace{z^2,\cdots,z^2}_{L},
\underbrace{\frac{1-rz}{1-r/z},\cdots,\frac{1-rz}{1-r/z}}_{M},
\underbrace{1,\cdots,1}_{R}\right),~~~(L+M+R=2N).
\end{eqnarray}
The scalar function $\varphi(z)$ in (\ref{def:K-matrix}) is
\begin{eqnarray}
\varphi(z)=\left\{\begin{array}{cc}
\varphi^{[1]}(z)& (L=M=0, R>0)\\
\varphi^{[2]}(z)& (L=0, M,R>0)\\
\varphi^{[3]}(z)& (L,M,R>0)
\end{array}
\right.,
\label{def:varphi}
\end{eqnarray}
where $L+M+R=2N$.
Here we have set
\begin{eqnarray}
&&\varphi^{[1]}(z)=(1-qz^2)^{\frac{1-N}{2N}},
\\
&&\varphi^{[2]}(z)=\varphi^{[1]}(z)\times\left\{\begin{array}{cc}
(1-rz/q)^{\frac{1-2N}{2N}}& (M={\rm odd})\\
1&(M={\rm even})\end{array}\right.,
\\
&&\varphi^{[3]}(z)=
\varphi^{[1]}(z) 
\times\left\{
\begin{array}{cc}
(1-z/rq)^{\frac{1-2N}{2N}} & (L={\rm odd}, M={\rm odd})\\
\left\{(1-z/rq)(1-rqz)\right\}^{\frac{1-2N}{2N}}& (L={\rm odd}, M={\rm even})\\
(1-rz/q)^{\frac{1-2N}{2N}}& (L={\rm even}, M={\rm odd})\\
1 & (L={\rm even}, M={\rm even})
\end{array}\right..
\end{eqnarray}
\end{dfn}
The boundary $K$-matrix $K(z)$ satisfies the graded boundary Yang-Baxter equation in $V \otimes V$
\begin{eqnarray}
K_2(z_2)R_{2,1}(z_1z_2)K_1(z_1)R_{1,2}(z_1/z_2)=
R_{2,1}(z_1/z_2)K_1(z_1)R_{1,2}(z_1z_2)K_2(z_2).
\label{eqn:BYBE}
\end{eqnarray}
The boundary $K$-matrix $K(z)$ satisfies (i) initial condition $K(1)=1$, 
(ii) boundary unitary condition $K(z)K(z^{-1})=1$, and (iii) boundary crossing symmetry
$\displaystyle K^*(z)K^*(z^{-1})=\frac{(1-rq^2z)(1-z/rq^2)}{(1-rz)(1-z/r)}$ with
$\displaystyle {K^*_j}^i(z)=\sum_{k,l=1}^{2N}
R_{l,k}^{j,i}(z^2) K_k^l(z)(-1)^{[v_k]+[v_i][v_j]}$.
The boundary $K$-matrix $K(z)$ given in (\ref{def:K-matrix}) is general diagonal solution of
the boundary Yang-Baxter equation.

\subsection{
$U_q(\widehat{gl}(N|N))$-analog of half-infinite $t-J$ model}

We introduce monodromy matrix ${\cal T}(z)$ by
\begin{eqnarray}
{\cal T}(z)=R_{0,1}(z)R_{0,2}(z) \cdots R_{0,n}(z) \in {\rm End}(V_n \otimes \cdots \otimes V_2 \otimes V_1 \otimes V_0),
\end{eqnarray}
where $V_j$ are copies of $V$ and $n \in {\bf N}$.
We introduce the transfer matrix $T_B^{fin}(z)$ by
\begin{eqnarray}
T_B^{fin}(z)={\rm str}_{V_0}(K(z^{-1})^{st}{\cal T}(z^{-1})^{-1}K(z){\cal T}(z))
\in {\rm End}(V_n \otimes \cdots \otimes V_2 \otimes V_1).
\end{eqnarray}
The Hamiltonian of $U_q(\widehat{gl}(N|N))$-analog of finite $t-J$ model
is given by
\begin{eqnarray}
H_B^{fin}=\frac{d}{dz}T_B^{fin}(z)|_{z=1}=\sum_{j=1}^{n-1}h_{j,j+1}+\frac{1}{2}\frac{d}{dz}K_1(z)|_{z=1}+
\frac{{\rm str}_{V_0}(K_0(1)^{st}h_{0,S})}{{\rm str}_{V_0}(K_0(1)^{st})},
\label{def:fin-Hamiltonian}
\end{eqnarray}
where $h_{j,j+1}=P_{j,j+1}\frac{d}{dz}R_{j,j+1}(z)|_{z=1}$.
We set the Hamiltonian $H_B$ by taking the thermodynamic limit of $H_B^{fin}$ in (\ref{def:fin-Hamiltonian}).
\begin{eqnarray}
H_B=\lim_{n \to \infty}H_B^{fin}=
\sum_{j=1}^{\infty}h_{j,j+1}+\frac{1}{2}\frac{d}{dz}K_1(z)|_{z=1}.
\label{def:infinite-Hamiltonian}
\end{eqnarray}
The Hamiltonian $H_B$ acts on the half-infinite tensor product space $\cdots \otimes V_3 \otimes V_2 \otimes V_1$.
We study 
$U_q(\widehat{gl}(N|N))$-analog of half-infinite $t-J$ model defined by the Hamiltonian $H_B$ in (\ref{def:infinite-Hamiltonian}).

\section{Vertex operator approach}
\label{sec:3}
In this Section we give the formulation of the vertex operator approach.

\subsection{Transfer matrix}

We would like to diagonalize the Hamiltonian $H_B$ in (\ref{def:infinite-Hamiltonian}).
It is convenient to study the transfer matrix 
\begin{eqnarray}
\widetilde{T}_B(z)=\lim_{n \to \infty}T_B^{fin}(z),
\end{eqnarray}
including spectral parameter $z$.
The transfer matrix $\widetilde{T}_B(z)$ is given by infinite product of the $R$-matrix.
Hence it isn't free from difficulty of divergence.
Later we would like to give mathematical formulation of our problem that is free from difficulty of divergence.
Following the strategy summarized in \cite{JM, JKKKM, JMN},
we introduce the vertex operator $\widetilde{\Phi}_j(z)$ and the dual vertex operator $\widetilde{\Phi}_j^*(z)$,
which act on half-infinite tensor product space $\cdots \otimes V_3 \otimes V_2 \otimes V_1$,
as limit of the monodromy matrix ${\cal T}(z)$.
The matrix elements of the vertex operator $\widetilde{\Phi}_j(z)$ $(j=1,2,\cdots,2N)$ are given by 
\begin{eqnarray}
\left(\widetilde{\Phi}_j(z)\right)_{\cdots j_n,\cdots,j_2,j_1}^{\cdots k_n,\cdots,k_2,k_1}
=\lim_{n \to \infty}
\left({\cal T}(z)\right)_{j_n,\cdots,j_2,j_1,j_0}^{k_n,\cdots,k_2,k_1, k_0}~~~~~{\rm for}~j=j_0.
\end{eqnarray}
The matrix elements of the dual vertex operator $\widetilde{\Phi}_j^*(z)$ $(j=1,2,\cdots,2N)$ are given by 
\begin{eqnarray}
\left(\widetilde{\Phi}_j^*(z)\right)_{\cdots k_n,\cdots,k_2,k_1}^{\cdots j_n,\cdots,j_2,j_1}
=\lim_{n \to \infty}
\left({\cal T}(z)^{-1}\right)_{k_n,\cdots,k_2,k_1,k_0}^{j_n,\cdots,j_2,j_1,j_0}~~~~~{\rm for}~j=j_0.
\end{eqnarray}
We expect that the vertex operator $\widetilde{\Phi}_j(z)$ and the dual vertex operator $\widetilde{\Phi}_j^*(z)$
give rise to well-defined operators.
From heuristic arguments by the Yang-Baxter equation \cite{JM, JKKKM, JMN}, the vertex operator $\widetilde{\Phi}_j(z)$ is expected to satisfy the following commutation relation.
\begin{eqnarray}
\widetilde{\Phi}_{j_2}(z_2)
\widetilde{\Phi}_{j_1}(z_1)=\sum_{k_1,k_2=1}^{2N}R_{j_1,j_2}^{k_1,k_2}(z_1/z_2)\widetilde{\Phi}_{k_1}(z_1)\widetilde{\Phi}_{k_2}(z_2)(-1)^{[v_{j_1}][v_{j_2}]}.
\end{eqnarray}
Here $R_{j_1,j_2}^{k_1,k_2}(z)$ is the matrix element of 
the $R$-matrix given in (\ref{def:R-matrix}).
The transfer matrix $\widetilde{T}_B(z)$
is written by using the vertex operators $\widetilde{\Phi}_j(z)$ and $\widetilde{\Phi}_j^*(z)$ as follows.
\begin{eqnarray}
\widetilde{T}_B(z)=\sum_{j=1}^{2N}\widetilde{\Phi}_j^*(z^{-1})K_j^j(z)\widetilde{\Phi}_j(z)(-1)^{[v_j]}.
\end{eqnarray}
The Hamiltonian $H_B$ is given by $H_B=\frac{d}{dz}\widetilde{T}_B(z)|_{z=1}$.
It is better to diagonalize $\widetilde{T}_B(z)$ instead of $H_B$.
In order to diagonalize $\widetilde{T}_B(z)$, we follow the strategy called the vertex operator approach.

\subsection{Vertex operator approach}

Our useful tool is the vertex operator associated with the quantum superalgebra $U_q(\widehat{gl}(N|N))$.
We introduce the evaluation representation $V_z$ of the basic representation $V=\oplus_{j=1}^{2N}{\bf C} v_j$ 
for $U_q(\widehat{gl}(N|N))$.
In what follows we use standard notation of $q$-integer
\begin{eqnarray}
[n]_q=\frac{q^n-q^{-n}}{q-q^{-1}}.
\end{eqnarray}
Let $E_{i,j}$ be $2N \times 2N$ matrix satisfying $E_{i,j}v_k=\delta_{j,k}v_i$.
The evaluation module $V_z$ is
given by the Drinfeld generators as follows.
\begin{eqnarray}
&&H_m^i
=(-1)^{i+1}\frac{[m]_q}{m}q^{(-1)^{i}m}(q^{[v_{i+1}]}z)^m(E_{i,i}+E_{i+1,i+1})~~~(1\leq i \leq 2N-1),
\nonumber\\
&&
H_m^{2N}=\frac{[2m]_q}{m}z^m\left\{-q^m \sum_{l=1}^N E_{2l,2l}
+\sum_{l=1}^N (1-N+(l-1)(1-q^m))(E_{2l-1,2l-1}+E_{2l,2l})\right\},
\nonumber
\\
&&
H_0^i=(-1)^{i+1}(E_{i,i}+E_{i+1,i+1})~~~(1\leq i \leq 2N-1),~~~
H_0^{2N}=\sum_{k=1}^{2N}(-1)^{k+1}E_{k,k},\\
&&
X_m^{+,i}=(q^{[v_{i+1}]}z)^m E_{i,i+1},
~~~
X_m^{-,i}=(-1)^{i+1} (q^{[v_{i+1}]}z)^m E_{i+1,i}
~~~(1\leq i \leq 2N-1).\nonumber
\end{eqnarray}
Denote by $V^{*S}=\oplus_{j=1}^{2N} {\bf C} v_j^*$
the left dual module of $V$ defined by
\begin{eqnarray}
\langle a \cdot v^*|v \rangle=(-1)^{[a][v^*]}\langle v^*| S(a) \cdot 
v\rangle~~~(a \in U_q(\widehat{gl}(N|N)), v^* \in V^{*S}, v\in V),
\end{eqnarray}
where the ${\bf Z}_2$-grading of the basis is chosen to be $[v_j^*]=\frac{(-1)^j+1}{2}$.
Namely the representation on $V^{*S}$ are given by
$\pi_{V^{*S}}(a)=\pi_{V}(S(a))^{st}$
where $\pi_V$ denotes action of module $V$.
The evaluation module $V_z^{* S}$ is
given by the Drinfeld generators as follows.
\begin{eqnarray}
&&H_m^i
=(-1)^{i}\frac{[m]_q}{m}q^{(-1)^{(i+1)}m}(q^{-[v_{i+1}]}z)^m(E_{i,i}+E_{i+1,i+1})~~~(1\leq i \leq 2N-1),
\nonumber\\
&&
H_m^{2N}=-\frac{[2m]_q}{m}z^m\left\{-q^{-m} \sum_{l=1}^N E_{2l,2l}
+\sum_{l=1}^N (1-N+(l-1)(1-q^{-m}))(E_{2l-1,2l-1}+E_{2l,2l})\right\},
\nonumber
\\
&&
H_0^i=(-1)^{i}(E_{i,i}+E_{i+1,i+1})~~~(1\leq i \leq 2N-1),~~~
H_0^{2N}=\sum_{k=1}^{2N}(-1)^{k}E_{k,k},\\
&&
X_m^{+,i}=-(-1)^i q^{(-1)^i}(q^{-[v_{i+1}]}z)^m E_{i+1,i},
~~~
X_m^{-,i}=-q^{(-1)^{i+1}}(q^{-[v_{i+1}]}z)^m E_{i,i+1}
~~~(1\leq i \leq 2N-1).
\nonumber
\end{eqnarray}

\begin{dfn}~~~
\label{def:vo}
Let $V(\lambda)$ be the highest-weight $U_q(\widehat{gl}(N|N))$-module with highest-weight $\lambda$.
We define the vertex operator $\Phi(z)$ and the dual vertex operator $\Phi^*(z)$ as the intertwiner of $U_q(\widehat{gl}(N|N))$-module as follows.
\begin{eqnarray}
\Phi(z) : V(\lambda) \longrightarrow V(\mu) \otimes V_z,&&~~~\Phi(z) \cdot x=\Delta(x) \cdot \Phi(z),\\
\Phi^*(z) : V(\lambda) \longrightarrow V(\mu) \otimes V_z^{* S},&&~~~\Phi^*(z) \cdot x=\Delta(x) \cdot \Phi^*(z),.
\end{eqnarray}
\end{dfn}
We expand the vertex operators as follows.
\begin{eqnarray}
\Phi(z)=\sum_{j=1}^{2N}\Phi_j(z) \otimes v_j,~~~\Phi^*(z)=\sum_{j=1}^{2N}\Phi_j^*(z) \otimes v_j^*.
\end{eqnarray}
We set the ${\bf Z}_2$-grading of the vertex operators by $[\Phi(z)]=[\Phi^*(z)]=0$.
Hence we have the ${\bf Z}_2$-grading $[\Phi_j(z)]=[\Phi_j^*(z)]=[v_j]=[v_j^*]$ $(j=1,2,\cdots,2N)$.
The vertex operators are expected to satisfy
the following relations.
\begin{eqnarray}
&&\Phi_{j_2}(z_2)\Phi_{j_1}(z_1)
=\sum_{k_1,k_2=1}^{2N}R_{j_1,j_2}^{k_1,k_2}(z_1/z_2)\Phi_{k_1}(z_1)\Phi_{k_2}(z_2)(-1)^{[v_{j_1}][v_{j_2}]},\\
&& g \Phi_i(z)\Phi_j^*(z)=(-1)^{[v_i]}\delta_{i,j},~~~
g \sum_{j=1}^{2N}(-1)^{[v_j]}\Phi_j^*(z)\Phi_j(z)=1.\label{vo:inversion}
\end{eqnarray}
Here $R_{j_1,j_2}^{k_1,k_2}(z)$ is matrix element of the $R$-matrix given in (\ref{def:R-matrix}).
Here $g$ is a constant. 
We introduce the transfer matrix $T_B(z)$ as follows.
\begin{eqnarray}
T_B(z)=g\sum_{i,j=1}^{2N} \Phi_i^*(z^{-1})K_i^j(z)\Phi_j(z)(-1)^{[v_j]}.
\end{eqnarray}
Following the strategy proposed in \cite{JM,JKKKM}, we consider our problem upon the following identification.
\begin{eqnarray}
T_B(z)=\widetilde{T}_B(z),~~~\Phi_j(z)=\widetilde{\Phi}_j(z),~~~\Phi_j^*(z)=\widetilde{\Phi}_j^*(z).
\end{eqnarray}
We call studies based on this identification "vertex operator approach".
The point of using the vertex operators $\Phi_j(z), \Phi_j^*(z)$ is that
they are well-defined objects and are free from difficulty of divergence.

\section{Bosonization of vertex operator}
\label{sec:4}
In this Section we review bosonizations of the vertex operators associated with the quantum superalgebra $U_q(\widehat{gl}(N|N))$.
We give integral representations of the dual vertex operator.

\subsection{Quantum superalgebra $U_q(\widehat{gl}(N|N))$}

We introduce bosons $\{a_m^i, c_m^j, Q_{a}^i, Q_{c^j}|i=1,2,\cdots,2N, j=1,2,\cdots,N, m \in {\bf Z}\}$
satisfying the following commutation relations.
\begin{eqnarray}
&&~[a_m^i,a_n^j]=(-1)^{i+1}\delta_{i,j}\delta_{m+n,0}\frac{[m]_q^2}{m},~~[a_0^i,Q_{a^j}]=\delta_{i,j}~~~(1\leq i,j \leq 2N),
\\
&&~[c_m^i,c_n^j]=\delta_{i,j}\delta_{m+n,0}\frac{[m]_q^2}{m},~~[c_0^i,Q_{c^j}]=\delta_{i,j}~~~(1\leq i,j \leq N).
\end{eqnarray}
The remaining commutators vanish.
We use standard normal ordering $::$ given by
\begin{eqnarray}
&&
:a_m^i a_n^i:=\left\{\begin{array}{cc}
a_m^i a_n^i&(m \leq 0)\\
a_n^i a_m^i&(m>0)
\end{array}\right.~(1\leq i \leq 2N),\\
&&
:a_0^i Q_{a^i}:=
:Q_{a_i} a_0^i:=Q_{a^i} a_0^i~(1\leq i \leq 2N),\nonumber
\\
&&
:c_m^j c_n^j:=\left\{\begin{array}{cc}
c_m^j c_n^j&(m \leq 0)\\
c_n^j c_m^j&(m>0)
\end{array}\right.~(1\leq j \leq N),
\\
&&
:c_0^i Q_{c^i}:=
:Q_{c_i} c_0^i:=Q_{c^i} c_0^i~(1\leq i \leq N).\nonumber
\end{eqnarray}
We set auxiliary bosonic operators $A_m^j, Q_{A}^j$ $(j=1,2,\cdots, 2N)$ by
\begin{eqnarray}
&&A_m^j=(-1)^{j+1}(a_m^j+a_m^{j+1}),~~~Q_{A^j}=Q_{a^j}-Q_{a^{j+1}}~~~(1\leq j \leq 2N-1),\\
&&A_m^{2N}=\frac{q^m+q^{-m}}{2}\sum_{l=1}^{2N}(-1)^{l+1}a_m^l,~~~Q_{A^{2N}}=\sum_{l=1}^{2N}Q_{a^l}.
\end{eqnarray}
They satisfy the following commutation relations.
\begin{eqnarray}
[A_m^i,A_n^j]=\frac{[A_{i,j}m]_q[m]_q}{m}\delta_{m+n,0},~~~
[A_0^i,Q_{A^j}]=A_{i,j}\delta_{m+n,0}~~~(1\leq i,j \leq 2N).
\end{eqnarray}
Here $A_{i,j}$ is matrix element of the Cartan matrix given in Appendix \ref{app:1}.
We set the notation
\begin{eqnarray}
&&
A^j(z;\kappa)=Q_{A^j}+A_0^j {\rm log}z-\sum_{m \neq 0}\frac{A_m^j}{[m]_q}q^{\kappa|m|}z^{-n}
~~~(j=1,2,\cdots,2N),\\
&&c^j(z)=Q_{c^j}+c_0^j {\rm log}z-\sum_{m\neq 0}\frac{c_m^j}{[m]_q}z^{-m}~~~(j=1,2,\cdots,N),
\\
&&
A_\pm^j(z)=\pm (q-q^{-1})
\sum_{m>0}A^j_{\pm m}
z^{\mp m}\pm A_0^j{\rm log}q~~~(j=1,2,\cdots, 2N). 
\end{eqnarray}
We set 
the auxiliary bosonic operators $A_m^{*j}, Q_{A^{*j}}$ $(j=1,2,\cdots, 2N)$ by
\begin{eqnarray}
&&
A_m^{*1}=\frac{1}{2N}\{(2N-1)a_m^1-\sum_{l=2}^{2N}a_m^l\},
\nonumber\\
&&
A_m^{* j}=\frac{1}{N}\{(N-1)\sum_{l=1}^{j}a_m^l-\sum_{l=j+1}^{2N}a_m^l\}~~~(1\leq j \leq 2N-2),
\\
&&
A_m^{* 2N-1}=\frac{1}{2N}\{\sum_{l=1}^{2N-1}a_m^l-(2N-1)a_m^{2N}\},~~~
A_m^{* 2N}=\frac{1}{2N}\sum_{l=1}^{2N}a_m^l,
\nonumber\\
&&
Q_{A^{*1}}=\frac{1}{2N}\{(2N-1)Q_{a^1}-\sum_{l=2}^{2N}Q_{a^l}\}
\nonumber\\
&&
Q_{A^{*j}}=\frac{1}{N}\{(N-1)\sum_{l=1}^{j}Q_{a^l}-\sum_{l=j+1}^{2N}Q_{a^l}\}
~~~(1\leq j \leq 2N-2),
\\
&&
Q_{A^{* 2N-1}}=\frac{1}{2N}\{\sum_{l=1}^{2N-1}Q_{a^l}-(2N-1)Q_{a^{2N}}\},~~~
Q_{A^{* 2N}}=\frac{1}{2N}\sum_{l=1}^{2N}Q_{a^l}.\nonumber
\end{eqnarray}
They satisfy the following commutation relations.
\begin{eqnarray}
&&[A_m^{*i},A_n^{*j}]=(\bar{A}^{-1})_{i,j}\frac{[m]_q^2}{m}\delta_{m+n,0}~~~(1\leq i,j \leq 2N),
\end{eqnarray}
where $\bar{A}=(A_{i,j})_{1\leq i,j \leq 2N}$
is the Cartan matrix given in Appendix \ref{app:1}.
They satisfy
\begin{eqnarray}
&&~[A_m^i,A_n^{*j}]=\delta_{m+n,0}\delta_{i,j}\frac{[m]_q^2}{m}~~~(1\leq i,j \leq 2N-1),
\nonumber\\
&&~[A_m^i,A_n^{* 2N}]=[A_m^{2N}, A_n^{* i}]=0~~~(1\leq i \leq 2N-1),
\nonumber\\
&&~[A_m^{2N},A_n^{* 2N}]=\frac{1}{2}\frac{[2m]_q[m]_q}{m}\delta_{m+n,0}.
\nonumber
\end{eqnarray}
For instance, we have
\begin{eqnarray}
&&[A_m^{*1},A_n^{*1}]=\frac{N-1}{N}\frac{[m]_q^2}{m}\delta_{m+n,0},
~~~[A_m^{*2N},A_n^{*2N}]=0,
\nonumber\\
&&[A_m^{*1},A_n^{*2N}]=\frac{1}{2N} \frac{[m]_q^2}{m}\delta_{m+n,0},
~~~[A_m^{* 2N-1},A_n^{* 2N-1}]=-\frac{N-1}{N}\frac{[m]_q^2}{m}\delta_{m+n,0},\nonumber
\\
&&[A_m^{* 2N-1},A_n^{*2N}]=\frac{1}{2N} \frac{[m]_q^2}{m}\delta_{m+n,0}.
\end{eqnarray}
We introduce the $q$-difference operator $\partial_z$ given by
\begin{eqnarray}
\partial_z f(z)=\frac{f(qz)-f(q^{-1}z)}{(q-q^{-1})z}.
\end{eqnarray}

\begin{thm}~\cite{Z}~~~
The Drinfeld generators of $U_q(\widehat{gl}(N|N))$ at Level-1 are 
realized as follows.
\begin{eqnarray}
&&c=1,~~~H_m^i=A_m^i~~~(m \in {\bf Z}, i=1,2,\cdots, 2N),~~~H_m^0=c-\sum_{j=1}^{2N-1}A_m^j,
\nonumber\\
&&
X^{\pm, i}(z)=:e^{\pm A^i(z;-\frac{1}{2})}Y^{\pm, i}(z)F^{\pm,i}:~~~(i=1,2,\cdots,2N-1),
\end{eqnarray}
\begin{eqnarray}
d=-\sum_{m>0}\frac{m^2}{[m]_q^2}\left\{\sum_{i=1}^{2N-1}A_{-m}^i A_m^{*i}+\frac{2}{q^m+q^{-m}}A_{-m}^{2N}A_m^{*2N}+\sum_{j=1}^N c_{-m}^j c_m^j\right\}
\nonumber\\
-\frac{1}{2}\left\{\sum_{i=1}^{2N} A_0^i A_0^{*i}+\sum_{j=1}^N c_0^j(c_0^j+1)\right\},\nonumber
\end{eqnarray}
where we have set
\begin{eqnarray}
&&
Y^{+,2j}(z)=Y^{-,2j-1}(z)=:\partial_z e^{c^j(z)}:~~~(1\leq j \leq N),
\nonumber\\
&&
Y^{-,2j}(z)=-Y^{+,2j-1}(z)=-:e^{c^j(z)}:~~~(1\leq j \leq N),
\nonumber\\
&&
F^{\pm, 2j-1}=\prod_{l=1}^{j-1}e^{\pm \sqrt{-1}a_0^{2l-1}},~~~
F^{\pm, 2j}=\prod_{l=1}^je^{\mp \sqrt{-1}\pi a_0^{2l}}~~~(1\leq j\leq N).
\end{eqnarray}
\end{thm}

We introduce the Fock module.
The vacuum vector $|0\rangle \neq 0$ is defined by
\begin{eqnarray}
a_m^i|0\rangle=c_m^j|0\rangle=0~~~(1\leq i \leq 2N, 1\leq j \leq N, m \geq 0).
\end{eqnarray}
For
$\lambda^a=(\lambda_1^a,\lambda_2^a,\cdots,\lambda_{2N}^a)\in {\bf C}^{2N}$ and $\lambda^c=(\lambda_1^c,\lambda_2^c,\cdots,\lambda_N^c)\in {\bf C}^N$,
we set
\begin{eqnarray}
|\lambda^a;\lambda^c\rangle=|
\lambda_1^a,\cdots,\lambda_{2N}^a;\lambda_1^c,\cdots,\lambda_N^c\rangle=e^{\sum_{i=1}^{2N}\lambda_i Q_{a^i}+\sum_{j=1}^N\lambda_j^c Q_{c^j}}|0\rangle.
\end{eqnarray}
Denote by ${F}_{\lambda_1^a,\cdots,\lambda_{2N}^a;\lambda_1^c,\cdots,\lambda_N^c}$
the Fock space generated by $\{a_{-m}^i, c_{-m}^j|m>0, i=1,2,\cdots,2N, j=1,2,\cdots,N\}$
over the vector $|\lambda^a;\lambda^c \rangle$.
Action of the bosonization of $U_q(\widehat{gl}(N|N))$ on 
${F}_{\lambda_1^a,\cdots,\lambda_{2N}^a;\lambda_1^c,\cdots,\lambda_N^c}$
is not closed.
We introduce the space
${\cal F}_{\lambda^a; \lambda^c}$ by
\begin{eqnarray}
{\cal F}_{\lambda^a; \lambda^c}=\bigoplus_{j_1,j_2,\cdots,j_{2N-1}\in {\bf Z}}
{F}_{\lambda_1^a+j_1,\lambda_2^a-j_1+j_2,\cdots,\lambda^a_{2N}-j_{2N-1};\lambda_1^c+j_1-j_2,\lambda_2^c+j_3-j_4,\cdots,\lambda_N^c+j_{2N-1}}.
\end{eqnarray}
Action of the bosonization for $U_q(\widehat{gl}(N|N))$ on ${\cal F}_{\lambda^a;\lambda^c}$ is closed.
\begin{eqnarray}
U_q(\widehat{gl}(N|N)){\cal F}_{\lambda^a;\lambda^c}
={\cal F}_{\lambda^a;\lambda^c}.
\end{eqnarray}
To obtain highest-weight vector, we impose the condition
\begin{eqnarray}
e_i|\lambda^a;\lambda^c \rangle=0~~~(i=0,1,2,\cdots,2N-1).
\end{eqnarray}
We have the following sufficient and necessary condition.
\begin{eqnarray}
&&
\lambda_1^a+\lambda_2^a+\lambda_1^c=0,1,~~
\lambda_{2j-1}^a+\lambda_{2j}^a+\lambda_j^c=0~~(2\leq j \leq N),\nonumber
\\
&&
\lambda_{2j}^a+\lambda_{2j+1}^a+\lambda_j^c=0~~(1\leq j \leq N-1).
\end{eqnarray}
For $\lambda_1^a+\lambda_2^a+\lambda_1^c=0$,
we have $\lambda_{2j-1}^a=\lambda_1^a$ $(2\leq j \leq N)$, 
$\lambda_j^c=-(\lambda_1^a+\lambda_{2j}^a)$ $(1\leq j \leq N)$.
For free parameter $\lambda_1^a=\beta_1, \lambda_{2j}^a=\beta_{2j} \in {\bf C}$ $(1\leq j \leq N)$, we have
the following highest-weight vector.
\begin{eqnarray}
|(1-(\beta_1+\beta_{2N}))\Lambda_0+\sum_{j=1}^{N-1}(\beta_1+\beta_{2j})(\Lambda_{2j-1}-\Lambda_{2j})+
(\beta_1+\beta_{2N})\Lambda_{2N-1}+(N\beta_1-\sum_{j=1}^N \beta_{2j})\Lambda_{2N}\rangle.
\label{def:lambda(1)}
\end{eqnarray}
As the special cases we have
\begin{eqnarray}
&&|\Lambda_0\rangle=\left|0,\cdots,0;0,\cdots,0\right.\rangle=|0\rangle,\\
&&|\Lambda_{2N-1}\rangle=\left|\frac{1}{2N},-\frac{1}{2N},\frac{1}{2N},-\frac{1}{2N},\cdots,\frac{1}{2N},-\frac{1}{2N}
\right.;
0,0,\cdots, 0,0,-1\rangle.
\end{eqnarray}
For $\lambda_1^a+\lambda_2^a+\lambda_1^c=1$,
we have $\lambda_{2j-1}^a=\lambda_1^a$ $(2\leq j \leq N)$, 
$\lambda_j^c=-(\lambda_1^a+\lambda_{2j}^a)+1$ $(1\leq j \leq N)$.
For free parameter $\lambda_1^a=\beta_1, \lambda_{2j}^a=\beta_{2j} \in {\bf C}$ $(1\leq j \leq N)$, we have
\begin{eqnarray}
&&|(1-(\beta_1+\beta_{2N}))\Lambda_0+(\beta_1+\beta_2)\Lambda_1+(1-(\beta_1+\beta_2))\Lambda_{2N-1}
\label{def:lambda(2)}
\\
&&
+\sum_{j=2}^{N-1}((\beta_1+\beta_{2j})-1)(\Lambda_{2j-1}-\Lambda_{2j})+
(\beta_1+\beta_{2N})\Lambda_{2N-1}
+(N\beta_1-\sum_{j=1}^N \beta_{2j}-N+1)\Lambda_{2N}\rangle.
\nonumber
\end{eqnarray}
As the special cases we have
\begin{eqnarray}
&&|\Lambda_1\rangle=\left|\frac{2N-1}{2N},\frac{1}{2N},-\frac{1}{2N},\frac{1}{2N}\cdots,
-\frac{1}{2N},\frac{1}{2N}
;0,0,0,\cdots,0\right.\rangle,\\
&&|\Lambda_2\rangle=
\left|\frac{N-1}{N},\frac{1-N}{N},-\frac{1}{N},\frac{1}{N},\cdots,-\frac{1}{N},\frac{1}{N}
\right.;
1,0,0, \cdots,0\rangle.
\end{eqnarray}
We are interested in the {\it irreducible} highest-weight module.
We have obtained bosonizations in the Fock space ${\cal F}_{\lambda^a;\lambda^c}$.
The module ${\cal F}_{\lambda^a;\lambda^c}$ is not irreducible in general.
To obtain irreducible representation, 
we introduce a pair of fermionic operators $\eta(z), \xi(z)$.
\begin{eqnarray}
\eta^i(z)=\sum_{n \in {\bf Z}}\eta_n^i z^{-n-1}=:e^{c^i(z)}:,~~~
\xi^i(z)=\sum_{n \in {\bf Z}}\xi_n^i z^{-n}=:e^{-c^i(z)}:~~~
(i=1,2,\cdots,N).
\end{eqnarray}
The Fourier components $\xi_m^j=\oint \frac{dw}{2\pi \sqrt{-1}}w^{m-1}\xi^j(w)$ and 
$\eta_m^j=\oint \frac{dw}{2\pi \sqrt{-1}}w^{m-1}\eta^j(w)$ are well-defined on the Fock space for $\lambda_c^j \in {\bf Z}$.
Hence we assume $\lambda_1^c,\lambda_2^c,\cdots,\lambda_N^c \in {\bf Z}$ in what follows.
They satisfy anti-commutation relations
\begin{eqnarray}
\xi_m^i \xi_n^i+\xi_n^i\xi_m^i=\eta_m^i\eta_n^i+\eta_n^i \eta_m^i=0,~~~
\xi_m^i \eta_n^i+\eta_n^i \xi_m^i=\delta_{m+n,0},~~~(i=1,2,\cdots,N).
\end{eqnarray}
The products $\eta_0^j \xi_0^j$ and $\xi_0^j \eta_0^j$ are orthogonal projection operators.
\begin{eqnarray}
\eta_0^j \xi_0^j+\xi_0^j \eta_0^j=1,~~~\eta_0^j \xi_0^j\cdot \xi_0^j \eta_0^j=
\xi_0^j \eta_0^j \cdot \eta_0^j \xi_0^j=0~~~(j=1,2,\cdots,N).
\end{eqnarray}
Hence we have direct sum decomposition
\begin{eqnarray}
{\cal F}_{\lambda^a;\lambda^c}=\eta_0^j \xi_0^j \cdot {\cal F}_{\lambda^a;\lambda^c} \oplus \xi_0^j \eta_0^j \cdot {\cal F}_{\lambda^a;\lambda^c}~~~(j=1,2,\cdots,N).
\end{eqnarray}
They commute with each other.
\begin{eqnarray}
[\xi_m^i,\xi_n^j]=[\eta_m^i,\eta_n^j]=[\xi_m^i,\eta_n^j]=0~~~(1\leq i \neq j \leq N).
\end{eqnarray}
We note that
\begin{eqnarray}
&&
\eta_0^j|\lambda^a;\lambda^c \rangle=0,~~~\xi_0^j|\lambda^a;\lambda^c \rangle \neq 0~~~(\lambda_j^c=0,1,2,\cdots),
\nonumber\\
&&
\eta_0^j|\lambda^a;\lambda^c \rangle\neq 0,~~~\xi_0^j|\lambda^a;\lambda^c \rangle=0~~~(\lambda_j^c=-1,-2,-3,\cdots).
\label{condition:annihilation}
\end{eqnarray}
We set the abbreviation $\zeta_\epsilon^j$ by
\begin{eqnarray}
\zeta_\epsilon^{j}=\left\{\begin{array}{cc}
\xi_0^j& (\epsilon=+)\\
\eta_0^j& (\epsilon=-)
\end{array}\right.~~~(j=1,2,\cdots,N).
\end{eqnarray}
We introduce the projection operator $Pr$ by
\begin{eqnarray}
Pr=\prod_{j=1}^N
\zeta_{\epsilon_j}^j \zeta_{-\epsilon_j}^j
~~~~~((\epsilon_1,\epsilon_2,\cdots,\epsilon_N)\in J(\lambda^c)),
\label{def:projection}
\end{eqnarray}
where we have used
\begin{eqnarray}
J(\lambda^c)=\left\{(\epsilon_1,\epsilon_2,\cdots,\epsilon_N)\left|\epsilon_j=\left\{\begin{array}{cc}
+ & (\lambda_j^c=-1,-2,\cdots)\\
- & (\lambda_j^c=0,1,2,\cdots)
\end{array}
\right.~~~(j=1,2,\cdots,N)\right.\right\}.
\end{eqnarray}
The projection operator $Pr$ commutes with every element of $U_q(\widehat{gl}(N|N))$, and satisfies $Pr^2=Pr$.
For general $N=1,2,3,\cdots$, 
we have the following direct sum of decomposition.
\begin{eqnarray}
{\cal F}_{\lambda^a;\lambda^c}=\bigoplus_{\epsilon_1,\epsilon_2,\cdots,\epsilon_N=\pm}
\prod_{j=1}^N 
\zeta_{\epsilon_j}^j\zeta_{-\epsilon_j}^j
{\cal F}_{\lambda^a;\lambda^c}.
\label{decomposition}
\end{eqnarray}
From calculation of character for $N=1,2$ in \cite{YZ2, YZ3},
we expect that the projection operator $Pr$ is a map onto an irreducible highest-weight module
$V(\lambda)$ with the highest-weight $\lambda$
given by (\ref{def:lambda(1)}) and (\ref{def:lambda(2)}).
\begin{eqnarray}
V(\lambda)=Pr \cdot {\cal F}_{\lambda^a;\lambda^c}.
\end{eqnarray}
The module $V(\lambda)=Pr \cdot {\cal F}_{\lambda^a;\lambda^c}$ is one of the sub-modules of the decomposition (\ref{decomposition}).

\subsection{Vertex operator}

In this Section we give
bosonizations of the vertex operators that intertwine irreducible highest-weight modules,
and derive integral representation of the dual vertex operator.
First, we give
the vertex operators $\phi(z), \phi^*(z)$ between
the Fock spaces ${\cal F}_{\lambda^a;\lambda^c}$.
We set the vertex operator $\phi(z)$ and the dual vertex operator $\phi^*(z)$ as the intertwiner of $U_q(\widehat{gl}(N|N))$-module as follows.
\begin{eqnarray}
\phi(z) : {\cal F}_{\lambda^a;\lambda^c} \longrightarrow 
{\cal F}_{\mu^a;\mu^c} \otimes V_z,&&~~~\phi(z) \cdot x=\Delta(x) \cdot \phi(z),
\label{def:vo2}\\
\phi^*(z) : {\cal F}_{\lambda^a;\lambda^c} \longrightarrow {\cal F}_{\mu^a;\mu^c} \otimes V_z^{* S},&&~~~\phi^*(z) \cdot x=\Delta(x) \cdot \phi^*(z).
\label{def:dual-vo2}
\end{eqnarray}
We expand the vertex operators as follows.
\begin{eqnarray}
\phi(z)=\sum_{j=1}^{2N}\phi_j(z) \otimes v_j,~~~\phi^*(z)=\sum_{j=1}^{2N}\phi_j^*(z) \otimes v_j^*.
\end{eqnarray}
We introduce the notation
\begin{eqnarray}
&&A^{* i}(z;\kappa)=Q_{A^{* i}}+A_0^{*i}{\rm log}z-\sum_{m \neq 0}\frac{A_m^{* i}}{[m]_q}q^{\kappa|m|}z^{-m}~~~(i=1,2,\cdots,2N-1),\nonumber
\\
&&B^1(z;\kappa)=Q_{A^{* 2N}}+A_0^{* 2N}{\rm log}z+2(N-1)\sum_{m \neq 0}\frac{A_m^{* 2N}}{[m]_q}q^{\kappa|m|}z^{-m},\\
&&B^{2N}(z;\kappa)=Q_{A^{* 2N}}+A_0^{* 2N}{\rm log}z-2N \sum_{m \neq 0}
\frac{A_m^{* 2N}}{[m]_q}q^{\kappa|m|}z^{-m},\nonumber
\end{eqnarray}
and the Fermi number operator $N_f=\sum_{j=1}^N a_0^{2j}$.

\begin{thm}~\cite{Z, YZ2, YZ3}~~~The vertex operators $\phi_j(z)$ and $\phi_j^*(z)$ $(j=1,2,\cdots,2N)$ in (\ref{def:vo2}) and (\ref{def:dual-vo2})
are realized as follows.
\begin{eqnarray}
&&
\phi_{2N}(z)=:e^{-A^{* 2N-1}(qz;\frac{1}{2})+B^{2N}(q^2z;\frac{1}{2})+c^N(qz)}:e^{-\sqrt{-1}\pi N_f},
\nonumber
\\
&&
(-1)^j\phi_j(z)=[\phi_{j+1}(z),f_j]_{q^{(-1)^j}}~~~(j=1,2,\cdots,2N-1),\\
&& \phi_1^*(z)=:e^{A^{*,1}(qz;\frac{1}{2})+B^1(qz;\frac{1}{2})}:e^{\sqrt{-1}\pi N_f}(qz)^{1-\lambda_1^a},
\nonumber\\
&&q^{(-1)^{j+1}}\phi_{j+1}^*(z)=[\phi_j^*(z),f_j]_{q^{(-1)^{j+1}}}~~~(j=1,2,\cdots,2N-1).
\end{eqnarray} 
Here the Chevalley generator $f_j$ is realized as follows.
\begin{eqnarray}
f_j=X_0^{-,j}=\oint \frac{dw}{2\pi \sqrt{-1}}X_j^-(w)~~~(j=1,2,\cdots,2N-1). 
\end{eqnarray}
\end{thm}

Next, we focus our attention to
the vertex operators $\Phi_j(z)$ between the irreducible highest-weight module $V(\lambda)$,
given in Definition \ref{def:vo} :
$\Phi(z) : V(\lambda) \longrightarrow V(\mu) \otimes V_z$,
$\Phi^*(z) : V(\lambda) \longrightarrow V(\mu) \otimes V_z^{* S}$.
We expect
\begin{eqnarray}
\Phi(z)=Pr \cdot \phi(z) \cdot Pr,~~~
\Phi^*(z)=Pr \cdot \phi^*(z) \cdot Pr,
\end{eqnarray}
where $Pr$ is the projection operator in (\ref{def:projection}).
The vertex operators $\phi(z)$ and $\phi^*(z)$ commute with $Pr$.
Hence we have
$\Phi(z)=Pr \cdot \phi(z)=\phi(z) \cdot Pr$ and
$\Phi^*(z)=Pr \cdot \phi^*(z)=\phi^*(z) \cdot Pr$.

In what follows we give integral representations of $\phi^*_j(z)$ which are convenient for construction of the boundary state.
We introduce the notation
\begin{eqnarray}
X_\epsilon^{-,2j-1}(z)=:e^{-H^{2j-1}(z;\frac{1}{2})-c^j(q^\epsilon z)}:F^{-,2j-1}.
\end{eqnarray}
Then we have $X^{-,2j-1}(z)=\frac{1}{(q-q^{-1})z}(X_+^{-,2j-1}(z)-X_-^{-,2j-1}(z))$.
Using the formulae of 
the normal orderings in Appendix \ref{app:3}, we have the following integral representations.

\begin{thm}~~~The dual vertex operators $\phi_j^*(z)$ $(j=1,2,\cdots,2N)$ 
have the following integral representations.
\begin{eqnarray}
\phi_{2j+1}^*(z)&=&
(q-q^{-1})^{j}
\sum_{\epsilon_1,\epsilon_3,\epsilon_5,\cdots,\epsilon_{2j-1}=\pm}
\prod_{s=1}^j\epsilon_{2s-1}
\oint_C \prod_{s=1}^{2j}\frac{dw_s}{2\pi \sqrt{-1}w_s}
\frac{qw_{2j}}{qz(1-qw_1/z)(1-qz/w_1)}\nonumber\\
&\times&\prod_{s=1}^{j-1}
\frac{1}{(1-qw_{2s+1}/w_{2s})(1-qw_{2s}/w_{2s+1})}
\prod_{s=1}^j
\frac{1}{(1-q^{\epsilon_{2s-1}}w_{2s-1}/w_{2s})}\nonumber\\
&\times&:\phi_1^*(z)\prod_{s=1}^{j} X_{\epsilon_{2s-1}}^{-,2s-1}(qw_{2s-1})\prod_{s=1}^{j}X^{-,2s}(qw_{2s}):~~~(j=1,2,\cdots,N-1),
\end{eqnarray}
where we take the integration contour $C$ to be simple closed curve that encircles $w_s=0, qw_{s-1}$
but not $q^{-1}w_{s-1}$ for $s=1,2,\cdots,2j$.
\begin{eqnarray}
\phi_{2j+2}^*(z)&=&
(q-q^{-1})^j
\sum_{\epsilon_1,\epsilon_3,\epsilon_5,\cdots,\epsilon_{2j+1}=\pm}\prod_{s=1}^{j+1} \epsilon_{2s-1}
\oint_C \prod_{s=1}^{2j+1}
\frac{dw_s}{2\pi \sqrt{-1}w_s}
\frac{1}{qz(1-qw_1/z)(1-qz/w_1)}\nonumber\\
&\times&\prod_{s=1}^{j}
\frac{1}{(1-qw_{2s+1}/w_{2s})(1-qw_{2s}/w_{2s+1})}
\prod_{s=1}^{j}\frac{1}{(1-q^{\epsilon_{2s-1}}w_{2s-1}/w_{2s})}\nonumber
\\
&\times&:\phi_1^*(z)\prod_{s=1}^{j+1} X_{\epsilon_{2s-1}}^{-,2s-1}(qw_{2s-1})
\prod_{s=1}^{j}X^{-,2s}(qw_{2s}):~~~(j=1,2,\cdots,N-1),
\end{eqnarray}
where we take the integration contour $C$ to be simple closed curve that encircles $w_s=0, qw_{s-1}$
but not $q^{-1}w_{s-1}$ for $s=1,2,\cdots,2j+1$.
\end{thm}

\section{Boundary state}
\label{sec:5}
In this Section we give a bosonization of the boundary state ${_B\langle 0|}\neq 0$ satisfying
\begin{eqnarray}
{_B\langle 0|}T_B(z)={_B\langle 0|}.\nonumber
\end{eqnarray}
The construction of the boundary state is main result of this paper.

\subsection{Bosonization}

For $\lambda^a=(\lambda_1^a,\lambda_2^a,\cdots,\lambda_{2N}^a)\in {\bf C}^{2N}$ and $\lambda^c=(\lambda_1^c,\lambda_2^c,\cdots,\lambda_N^c)\in {\bf C}^N$,
we set the vector 
$\langle \lambda^a;\lambda^c |$ by
\begin{eqnarray}
\langle \lambda^a;\lambda^c |=\langle 0|e^{-\sum_{i=1}^{2N} \lambda_i^a Q_{a^{i}}
-\sum_{j=1}^N \lambda_j^c Q_{c^j}},
\label{def:lambda}
\end{eqnarray}
where the vector $\langle 0|\neq 0$ satisfies
\begin{eqnarray}
\langle 0|a_{-m}^i=\langle 0|c_{-m}^j=0~~~(m \geq 0, i=1,2,\cdots,2N, j=1,2,\cdots,N).
\end{eqnarray}
We have introduce the space
${\cal F}^*_{\lambda^a; \lambda^c}$ by
\begin{eqnarray}
{\cal F}^*_{\lambda^a; \lambda^c}=\bigoplus_{j_1,j_2,\cdots,j_{2N-1}\in {\bf Z}}
{F}^*_{\lambda_1^a+j_1,\lambda_2^a-j_1+j_2,\cdots,\lambda^a_{2N}-j_{2N-1};\lambda_1^c+j_1-j_2,\lambda_2^c+j_3-j_4,\cdots,\lambda_N^c+j_{2N-1}}.
\end{eqnarray}
Denote by ${F}^*_{\lambda_1^a,\cdots,\lambda_{2N}^a;\lambda_1^c,\cdots,\lambda_N^c}$
the dual Fock space generated by $\{a_{m}^i, c_{m}^j|m>0, i=1,2,\cdots,2N, j=1,2,\cdots,N\}$
over the vector $\langle \lambda^a;\lambda^c|$.
Following arguments in the previous section, we expect the following identification
between the restricted dual module $V^*(\lambda)$ and the Fock module. 
\begin{eqnarray}
V^*(\lambda)={\cal F}_{\lambda^a;\lambda^c}^{~*} \cdot Pr.
\end{eqnarray}
For instance we have the highest-weight vectors $\langle \Lambda_i| \in V^*(\Lambda_i)$ as follows.
\begin{eqnarray}
&&\langle \Lambda_0|=\langle \left.0,\cdots,0;0,\cdots,0\right|,
\label{lambda2N-1}\\
&&\langle \Lambda_{2N-1}|=
\langle \frac{1}{2N},-\frac{1}{2N},\frac{1}{2N},-\frac{1}{2N},\cdots,\frac{1}{2N},-\frac{1}{2N};
0,0,\cdots, 0,0,-1|,
\label{lambda2N-1}\\
&&\langle \Lambda_1|
=\langle \frac{2N-1}{2N},\frac{1}{2N},-\frac{1}{2N},\frac{1}{2N}\cdots,
-\frac{1}{2N},\frac{1}{2N}
;0,0,0,\cdots,0|,
\label{lambda1}\\
&&\langle \Lambda_2|
=
\langle \frac{N-1}{N},\frac{1-N}{N},-\frac{1}{N},\frac{1}{N},\cdots,-\frac{1}{N},\frac{1}{N};
1,0,0, \cdots,0|.
\label{lambda2}
\end{eqnarray}
In this Section we focus our attention to
the integrable highest-weight module $V^*(\Lambda_0)$ for simplicity.
Later we summarize results associated with the integrable modules $V^*(\Lambda_{2N-1})$ 
in Appendix \ref{app:2}.
For $V^*(\Lambda_0)$ the projection operator $Pr$ is given by
$\displaystyle Pr=\prod_{j=1}^N \eta_0^j \prod_{j=1}^N \xi_0^j$.

\begin{dfn}~~~We define the bosonic operator $G$ by
\begin{eqnarray}
G&=&-\frac{1}{2}\sum_{m>0}
\frac{mq^{-2m}}{[m]_q^2}\left(\sum_{i=1}^{2N}(-1)^{i+1}(a_m^i)^2+\sum_{j=1}^N (c_m^j)^2\right)
+
\sum_{m>0}\left(\sum_{i=1}^{2N}\beta_m^i a_m^i+\sum_{j=1}^N \delta_m^j c_m^j\right).\nonumber\\
\label{def:G}
\end{eqnarray}
Here we have set
\begin{eqnarray}
\delta_m^j&=&\frac{q^{-m}}{[m]_q}(1-2(-1)^{\frac{m}{2}})\theta_m~~(j=1,2,\cdots,N),
\label{def:delta}
\\
\beta_m^i&=&\left\{\begin{array}{cc}
\beta_m^{[1], i}& (L=M=0, R>0)\\
\beta_m^{[2], i}& (L=0,M,R>0)\\
\beta_m^{[3], i}& (L,M,R>0)
\end{array}\right.~~(i=1,2,\cdots,2N),
\label{def:beta}
\end{eqnarray}
where we have set
\begin{eqnarray}
&&
\beta_m^{[1],2s-1}=\beta_m^{[1], 2s}=\frac{2(1-s)}{[m]_q}q^{-\frac{3}{2}m}\theta_m~~~(1\leq s \leq N),\\
&&
\beta_m^{[2],i}=\beta_m^{[1],i}-\frac{r^m q^{-\frac{3}{2}m}}{[m]_q}+\left\{\begin{array}{cc}
0 & (1\leq j \leq M)\\
\frac{\displaystyle r^m q^{-\frac{5}{2}m}}{\displaystyle [m]_q}& (M<i \leq 2N, M={\rm odd})\\
\frac{\displaystyle r^m q^{-\frac{3}{2}m}}{\displaystyle [m]_q}& (M<i \leq 2N, M={\rm even})
\end{array}\right.,\\
&&
\beta_m^{[3],i}=\beta_m^{[1],i}-\frac{(-1)^{\frac{m}{2}}2 q^{-\frac{3}{2}m}}{[m]_q}\theta_m+
\left\{\begin{array}{cc}
0& (1\leq i \leq L)\\
\frac{\displaystyle r^{-m} q^{-\frac{5}{2}m}}{\displaystyle [m]_q}&(L<i\leq L+M, L={\rm odd})\\
\frac{\displaystyle r^{-m} q^{-\frac{3}{2}}m}{\displaystyle [m]_q}& (L<i \leq L+M, L={\rm even})\\
\frac{\displaystyle r^{m}q^{\frac{1}{2}m}}{\displaystyle [m]_q}& (L+M<i \leq 2N, L={\rm odd}, M={\rm odd})\\
\frac{\displaystyle r^{m}q^{-\frac{1}{2}m}}{\displaystyle [m]_q}& (L+M<i \leq 2N, L={\rm odd}, M={\rm even})\\
\frac{\displaystyle r^{m}q^{-\frac{5}{2}m}}{\displaystyle [m]_q}& (L+M<i \leq 2N, L={\rm even}, M={\rm odd})\\
\frac{\displaystyle r^{m}q^{-\frac{3}{2}m}}{\displaystyle [m]_q}& (L+M<i \leq 2N, L={\rm even}, M={\rm even})
\end{array}\right..\nonumber\\
\end{eqnarray}
Here we have used
\begin{eqnarray}
\theta_m=\left\{\begin{array}{cc}
1& (m={\rm even})\\
0& (m={\rm odd})
\end{array}\right..
\end{eqnarray}
\end{dfn}
The following theorem is {\bf main result} of this paper.

\begin{thm}~~~\label{thm:main}
A bosonization of the boundary state ${_B \langle 0|} \in V^*(\Lambda_0)$ is realized as follows.
\begin{eqnarray}
{_B\langle 0|}=\langle \Lambda_0|e^G \cdot Pr.
\end{eqnarray}
Here $G$ is given in (\ref{def:G}), and $Pr$ is the projection operator.
\end{thm}

\section{Proof}
\label{sec:6}
In this Section we give a proof of the boundary state.
The following proposition gives sufficient condition of Theorem \ref{thm:main}.

\begin{prop}~~~A sufficient condition of ${_B\langle 0|}T_B(z)={_B\langle 0|}$ 
is given by
\begin{eqnarray}
\langle 0|e^G \phi_j^*(z^{-1})K_j^j(z)=\langle 0|e^G \phi_j^*(z)~~~(j=1,2,\cdots,2N).\label{eqn:sufficient}
\end{eqnarray}
\end{prop}
{\it Proof.}~~~Multiplying $\Phi_j^*(z)$ to ${_B\langle 0|}T_B(z)={_B\langle 0|}$ 
from the right and using the inversion relation of the vertex operators (\ref{vo:inversion}),
we have ${_B\langle 0|}\Phi_j^*(z^{-1})K_j^j(z)={_B\langle 0|}
\Phi_j^*(z)$. The projection operator $Pr$ commutes with the vertex operators, we have
$\langle 0| e^G \phi_j^*(z^{-1})K_j^j(z)\cdot Pr=\langle 0|e^G \phi_j^*(z)\cdot Pr$. 
Removing the projection operator $Pr$, we have the above sufficient condition.
{\bf Q.E.D.}

\subsection{Action of vertex operator}

In this Section we study the action of the vertex operator on the boundary state.

\begin{prop}~~~\label{action:G}
\begin{eqnarray}
e^G a_{-m}^j e^{-G}&=&a_{-m}^j-q^{-2m}a_m^j+(-1)^{j+1} \beta_m^j \frac{[m]_q^2}{m}~~~(m>0, 1\leq i \leq 2N),\\
e^G c_{-m}^j e^{-G}&=&c_{-m}^j-q^{-2m}c_m^j+\delta_m^j \frac{[m]_q^2}{m}~~~(m>0, 1\leq j \leq N),\\
e^G A_{-m}^j e^{-G}&=&A_{-m}^j-q^{-2m}A_m^j+\frac{[m]_q^2}{m}(\beta_m^j-\beta_m^{j+1})~~~(m>0, 1\leq i \leq 2N-1).
\end{eqnarray}
\end{prop}
{\it Proof.}~~~
We note that 
$$e^G Xe^{-G}=e^{ad(G)}(X)=\sum_{n=0}^\infty \frac{1}{n!}ad(G)^n(X).$$
Using $ad^2(G)(X)=[G,[G,X]]=0$ for $X=a_{m}^i, c_m^j$, 
we have 
$$e^{G}Xe^{-G}=X+[G,X].$$
Using the following relations, we get proposition.
\begin{eqnarray}
~[G,a_{-m}^i]&=&-q^{-2m}a_m^i+\frac{[m]_q^2}{m}(-1)^{i+1}\beta_m^i,
\\
~[G,c_{-m}^j]&=&-q^{-2m}c_m^j+\frac{[m]_q^2}{m}\delta_m^j,
\\
~[G,A_{-m}^i]&=&-q^{-2m}A_m^i+\frac{[m]_q^2}{m}(\beta_m^i-\beta_m^{i+1}).
\end{eqnarray}
{\bf Q.E.D.}
\\
We set notations.
\begin{eqnarray}
&&A_\pm^i(z;\kappa)=\mp \sum_{m=1}^\infty \frac{A_{\pm m}^i}{[m]_q}q^{\kappa m}z^{\mp m}~~~(i=1,2,\cdots,2N-1),\\
&&B_\pm(z;\kappa)=\mp \sum_{m=1}^\infty \frac{1}{[m]_q}\left(A_{\pm m}^{* 1}-2(N-1)A_{\pm m}^{* 2N}\right)q^{\kappa m}z^{\mp m},\\
&&c_\pm^j(z)=\mp \sum_{m=1}^\infty \frac{c_{\pm m}^j}{[m]_q}z^{\mp m}~~~(j=1,2,\cdots,N).
\end{eqnarray}
We have
$A^{*1}(z;\kappa)+B^1(z;\kappa)=Q_{a^1}+a_0^1{\rm log}z+B_+(z;\kappa)+B_-(z;\kappa)$.

\begin{prop}~~~
The actions of the basic operators are given as follows.
\begin{eqnarray}
\langle 0|e^G e^{-A_-^i(qw;\frac{1}{2})}
&=&
g_i(w)\langle 0|e^G e^{-A_+^i(q/w;\frac{1}{2})}~~~(i=1,2,\cdots,2N-1),\\
\langle 0|e^G e^{B_-(qz;\frac{1}{2})}
&=&
f_1(z) \langle 0|e^G e^{B_+(q/z;\frac{1}{2})},
\\
\langle 0|e^G e^{c_-^j(qw)}&=&
c_j(w)\langle 0|e^G e^{c_+^j(q/w)}~~~(j=1,2,\cdots,N),
\end{eqnarray}
where $c_j(w)$, $f_1(z)$, and $g_i(w)$ are given as follows.
\begin{eqnarray}
c_j(w)=1+w^2~~~(j=1,2,\cdots,N),\label{def:c}
\end{eqnarray}
\begin{eqnarray}
f_1(z)&=&\left\{\begin{array}{cc}
\varphi^{[1]}(z)& (L=M=0, R>0)\\
(1-rz)\varphi^{[2]}(z)& (L=0, M,R>0)\\
(1+z^2)\varphi^{[3]}(z)& (L,M,R>0)
\end{array}
\right.,
\label{def:f}
\end{eqnarray}
\begin{eqnarray}
g_i(w)&=&\left\{\begin{array}{cc}
g_i^{[1]}(w)& (L=M=0,R>0)\\
g_i^{[2]}(w)& (L=0,M,R>0)\\
g_i^{[3]}(w)& (L,M,R>0)
\end{array}
\right.~~~(i=1,2,\cdots,2N-1).
\label{def:g}
\end{eqnarray}
Here we have set
\begin{eqnarray}
g_i^{[1]}(w)&=&\left\{\begin{array}{cc}
1&(i={\rm odd})\\
(1-w^2)& (i={\rm even})
\end{array}\right.,\\
g_i^{[2]}(w)&=&g_i^{[1]}(w) \times \left\{\begin{array}{cc}
\frac{1}{(1-rw/q)}& (i=M, M={\rm odd})\\
\frac{1}{(1-rw)}& (i=M, M={\rm even})\\
1 & ({\rm otherwise})
\end{array}
\right.,\\
g_i^{[3]}(w)&=&g_i^{[1]}(w)\times
\left\{
\begin{array}{cc}
\frac{1}{(1-w/qr)}&~~~(i=L, L={\rm odd})
\\
\frac{1}{(1-w/r)}&~~~(i=L, L={\rm even})
\\
\frac{1}{(1-q^2r w)}&~~~(i=L+M, L={\rm odd}, M={\rm odd})
\\
\frac{1}{(1-qr w)}&~~~(i=L+M, L={\rm odd}, M={\rm even})
\\
\frac{1}{(1-rw/q)}&~~~(i=L+M, L={\rm even}, M={\rm odd})
\\
\frac{1}{(1-rw)}&~~~(i=L+M, L={\rm even}, M={\rm even})
\\
1 & ({\rm otherwise})
\end{array}
\right..
\end{eqnarray}
\end{prop}
{\it Proof.}~~~Using Proposition \ref{action:G}, we have
\begin{eqnarray}
&&
e^G e^{-A_-^j(qw;\frac{1}{2})}e^{-G}=g_j(w)e^{-A_-^j(qw;\frac{1}{2})}e^{-A_+^j(q/w;\frac{1}{2})},\\
&&
e^G e^{B_-(qz;\frac{1}{2})}e^{-G}=
f_1(z)e^{B_-(qz;\frac{1}{2})}
e^{B_+(q/z;\frac{1}{2})},\\
&&
e^G e^{-c_-^j(qw)} e^{-G}=c_j(w)
e^{-c_-^j(qw)}e^{-c_+^j(q/w)},
\end{eqnarray}
where
\begin{eqnarray}
g_j(w)&=&
\exp\left(-\sum_{m>0}\frac{[m]_q}{m}q^{\frac{3}{2}m}(\beta_m^j-\beta_m^{j+1})w^m\right)~~~
(1\leq j \leq 2N-1),\\
f_1(z)&=&\exp\left(\frac{N-1}{N}\sum_{m>0}\frac{1}{2m}q^m z^{2m}
+\sum_{m>0}\frac{[m]_q}{m}q^{\frac{3}{2}m}\beta_m^1 z^m \right.\nonumber\\
&&\left.
+\left(1-\frac{1}{2N}\right)\sum_{m>0}\frac{[m]_q}{m}q^{\frac{3}{2}m}
\left(\sum_{j=1}^{2N}(-1)^j \beta_m^j \right) z^m
\right),\\
c_j(w)&=&\exp\left(
-\sum_{m>0}\frac{1}{2m}w^{2m}+\sum_{m>0}\frac{[m]_q}{m}q^m \delta_m^j w^m
\right)~~~
(1\leq j \leq N).
\end{eqnarray}
Using (\ref{def:delta}) and (\ref{def:beta}), we have the explicit formulae of 
$g_j(w)$, $f_1(z)$, and $c_j(w)$.
Acting these operators to the vacuum vector $\langle 0|$, we obtain this proposition.
~~~{\bf Q.E.D.}
\\
We set
\begin{eqnarray}
&&
D(z,w)=(1-qzw)(1-qw/z)(1-qz/w)(1-q/zw),
\label{def:D}\\
&&
H_\epsilon^{2j-1}(w_1,w_2)=\frac{e^{-c_+^j(q^{1+\epsilon}w_1)-c_+^j(q^{1-\epsilon}/w_1)-A_+^{2j-1}(qw_1)-A_+^{2j-1}(q/w_1)}}
{(1-q^\epsilon w_1w_2)(1-q^\epsilon w_1/w_2)},
\label{def:H1}
\\
&&
H^{2j}(w_1,w_2)=\frac{e^{-A_+^{2j}(qw_1)-A_+^{2j}(q/w_1)+c_+^j(qw_1)+c_+^j(q/w_1)}}{D(w_1,w_2)},
\label{def:H2}
\\
&&
J_\epsilon^{2j-1}(w)=e^{-c_+^j(q^{1+\epsilon}w)-c_+^j(q^{1-\epsilon}/w)-A_+^{2j-1}(qw)-A_+^{2j-1}(q/w)},
\label{def:J1}
\\
&&
J^{2j}(w)=e^{-A_+^{2j}(qw)-A_+^{2j}(q/w)+c_+^j(qw)+c_+^j(q/w)}.
\label{def:J2}
\end{eqnarray}

\begin{prop}~~~
\begin{eqnarray}
&&
D(z,w)=D(z^{-1},w)=D(z,w^{-1})=D(z^{-1},w^{-1}),
\nonumber
\\
&&
H_\epsilon^{2j-1}(w_1,w_2)=q^{-2\epsilon}w_1^2 H_{-\epsilon}^{2j-1}(w_1^{-1},w_2),~~~
H_\epsilon^{2j-1}(w_1,w_2)=H_\epsilon^{2j-1}(w_1,w_2^{-1}),
\nonumber\\
&&
H^{2j}(w_1,w_2)=H^{2j}(w_1^{-1},w_2)=H^{2j}(w_1,w_2^{-1})=H^{2j}(w_1^{-1},w_2^{-1}),
\nonumber\\
&&J_\epsilon^{2j-1}(w)=J_{-\epsilon}^{2j-1}(w^{-1}),~~~
J^{2j}(w)=J^{2j}(w^{-1}).
\end{eqnarray}
\end{prop}

\begin{prop}~~~The following relations for $H_\epsilon^{2j-1}(w_1,w_2)~(j=1,2,\cdots,N, \epsilon=\pm)$ hold.
\begin{eqnarray}
&&
\oint_{{C}} \frac{dw_1}{w_1}\sum_{\epsilon=\pm}\epsilon q^{\epsilon}
\left(q^\epsilon w_1+\frac{1}{q^\epsilon w_1}\right)(1-qw_1w_2) H_\epsilon^{2j-1}(w_1,w_2)\nonumber\\
&&=
\oint_{{C}} \frac{dw_1}{w_1}q \left(qw_1+\frac{1}{qw_1}\right)(1-w_1^2)H_+^{2j-1}(w_1,w_2),\label{eqn:recursion1}
\end{eqnarray}
\begin{eqnarray}
&&
\oint_{{C}} \frac{dw_1}{w_1}
\sum_{\epsilon=\pm}\epsilon q^{\epsilon}
\left(q^\epsilon w_1+\frac{1}{q^\epsilon w_1}\right)
(1-sw_1)(1-qw_1w_2) H_\epsilon^{2j-1}(w_1,w_2)
\nonumber\\
&&=
(1-qsw_2)\oint_{{C}} \frac{dw_1}{w_1}
q \left(q w_1+\frac{1}{q w_1}\right)
(1-w_1^2)H_+^{2j-1}(w_1,w_2),
\label{eqn:recursion2}
\end{eqnarray}
\begin{eqnarray}
&&
\oint_{{C}} \frac{dw_1}{w_1}
\sum_{\epsilon=\pm}\epsilon q^{\epsilon}
\left(q^\epsilon w_1+\frac{1}{q^\epsilon w_1}\right)
\frac{(1-qw_1w_2)}{(1-s/w_1)} H_\epsilon^{2j-1}(w_1,w_2)\nonumber\\
&&=(1-qsw_2)\oint_{{C}} \frac{dw_1}{w_1}q
\left(q w_1+\frac{1}{q w_1}\right)
\frac{(1-w_1^2)}{(1-s/w_1)(1-sw_1)}H_+^{2j-1}(w_1,w_2).
\label{eqn:recursion3}
\end{eqnarray}
\begin{eqnarray}
&&
\oint_{{C}} \frac{dw_1}{w_1}\sum_{\epsilon=\pm}\epsilon q^{\epsilon}
\left(q^\epsilon w_1+\frac{1}{q^\epsilon w_1}\right)w_1(1-qw_1w_2) H_\epsilon^{2j-1}(w_1,w_2)\nonumber\\
&&=q^2 w_2
\oint_{{C}} \frac{dw_1}{w_1}\left(qw_1+\frac{1}{qw_1}\right)(1-w_1^2)H_+^{2j-1}(w_1,w_2).\label{eqn:recursion4}
\end{eqnarray}
Here the integration contour ${C}$ encircles $w_1=0$, $s$, $qw_2^{\pm 1}$ but not $s^{-1}$, $q^{-1} w_2^{\pm 1}$.
Note that
the integral
$\oint_{{C}} \frac{dw_1}{w_1}
\left(q w_1+\frac{1}{q w_1}\right)
\frac{(1-w_1^2)}{(1-s/w_1)(1-sw_1)}H_+^{2j-1}(w_1,w_2)$
is invariant under $w_2 \to w_2^{-1}$.
\end{prop}
{\it Proof.}~~~
We show the second relation (\ref{eqn:recursion2}).
We start from LHS :
\begin{eqnarray}
&&\oint_C \frac{dw_1}{w_1} q
\left(q w_1+\frac{1}{q w_1}\right)
(1-sw_1)(1-qw_1w_2) H_+^{2j-1}(w_1,w_2)\nonumber\\
&-&
\oint_C \frac{dw_1}{w_1} q^{-1}
\left(\frac{w_1}{q}+\frac{q}{w_1}\right)
(1-sw_1)(1-qw_1w_2) H_-^{2j-1}(w_1,w_2).
\nonumber
\end{eqnarray}
Changing the variable $w_1 \to w_1^{-1}$ in the second term and using $H_-^{2j-1}(w_1^{-1},w_2)=q^2 w_1^2 H_+^{2j-1}(w_1,w_2)$,
we have RHS after summing up the first and the second terms.
The relations (\ref{eqn:recursion3}) and (\ref{eqn:recursion4}) are obtained in the same way.
Upon the specialization $s=0$, (\ref{eqn:recursion1}) is obtained from both 
(\ref{eqn:recursion2}) and (\ref{eqn:recursion3}).~~~{\bf Q.E.D.}

\begin{prop}~~~
The following relations for $H^{2j}(w_1,w_2)~(j=1,2,\cdots,N)$ hold.
\begin{eqnarray}
&&\oint_C \frac{dw_1}{w_1}(1-sw_1)(1-q/w_1w_2)\frac{(1-w_1^2)}{(1+w_1^2)}H^{2j}(w_1,w_2)\nonumber\\
&=& -\frac{q}{2w_2}(1-sw_2/q)\times \oint_C \frac{dw_1}{w_1} \frac{(1-w_1^2)^2}{w_1(1+w_1^2)} H^{2j}(w_1,w_2),
\label{eqn:recursion5}
\end{eqnarray}
\begin{eqnarray}
\oint_C \frac{dw_1}{w_1}w_1(1-q/w_1w_2)\frac{(1-w_1^2)}{(1+w_1^2)}H^{2j}(w_1,w_2)
= -\frac{1}{2}\times \oint_C \frac{dw_1}{w_1} \frac{(1-w_1^2)^2}{w_1(1+w_1^2)} H^{2j}(w_1,w_2),
\label{eqn:recursion6}
\end{eqnarray}
\begin{eqnarray}
&&\oint_C \frac{dw_1}{w_1}\frac{(1-q/w_1w_2)}{(1-s/w_1)}\frac{(1-w_1^2)}{1+w_1^2}H^{2j}(w_1,w_2)\nonumber\\
&=& -\frac{q}{2w_2}(1-sw_2/q)\times \oint_C \frac{dw_1}{w_1} 
\frac{(1-w_1^2)^2}{w_1(1+w_1^2)} 
\frac{H^{2j}(w_1,w_2)}{
(1-sw_1)(1-s/w_1)}.
\label{eqn:recursion7}
\end{eqnarray}
Here the integration contour 
${C}$ encircles $w_1=0$, s, $\sqrt{-1}$, $q w_2^{\pm 1}$ but not $s^{-1}, -\sqrt{-1}$, $q^{-1} w_2^{\pm 1}$.
Note that
the integral
$\oint_C \frac{dw_1}{w_1} \frac{(1-w_1^2)^2}{w_1(1+w_1^2)} H^{2j}(w_1,w_2)$ is invariant under $w_2 \to w_2^{-1}$.
\end{prop}
{\it Proof.}~~~
We start from LHS. Taking into account of the relation
$\oint \frac{dw}{w}f(w)=\frac{1}{2}\oint \frac{dw}{w}(f(w)+f(w^{-1}))$ and using $H^{2j}(w_1^{-1},w_2)=H^{2j}(w_1,w_2)$,
we have
\begin{eqnarray}
\frac{1}{2}
\oint_C \frac{dw_1}{w_1}
\left\{
(1-sw_1)(1-q/w_1w_2)
-(1-s/w_1)(1-qw_1/w_2)
\right\}
\frac{1-w_1^2}{1+w_1^2}
H^{2j}(w_1,w_2)
.\nonumber
\end{eqnarray}
From $(1-sw_1)(1-q/w_1w_2)-(w_1 \leftrightarrow w_1^{-1})=\frac{q}{w_1w_2}(w_1^2-1)(1-sw_2/q)$,
the relation (\ref{eqn:recursion5}) is obtained.
The relations (\ref{eqn:recursion6}) and (\ref{eqn:recursion7}) are obtained in the same way.
~~~{\bf Q.E.D.}
\\
\\
We have the actions of the vertex operators as follows.

\begin{prop}~~~
The actions of the vertex operators $\phi_1^*(z), \phi_2^*(z), \cdots, \phi_{2N}^*(z)$
are given as follows.
\begin{eqnarray}
\langle \lambda^a;\lambda^c |e^G \phi_{2j}^*(z)
&=&(-1)(q-q^{-1})^{j-1}\prod_{s=1}^N e^{\sqrt{-1}\pi \lambda_{2s}^a}\prod_{s=1}^j\prod_{t=1}^s e^{-\sqrt{-1}\pi \lambda_{2t-1}^a}
\prod_{s=1}^{j-1}\prod_{t=1}^s e^{-\sqrt{-1}\pi \lambda_{2t}^a}
f_1(z) 
\nonumber\\
&\times&
\sum_{\epsilon_1,\epsilon_3,\cdots,\epsilon_{2j-1}=\pm}
\prod_{s=1}^j \epsilon_{2s-1} \cdot
q^{-\epsilon_{2j-1} (\lambda_j^c+1)}
\prod_{s=1}^{j-1}q^{-\lambda_s^c \epsilon_{2s-1}}
\prod_{s=1}^{2j-1}\oint_{{C}}
\frac{dw_{s}}{2\pi \sqrt{-1} w_s}\nonumber\\
&\times&
\prod_{s=1}^j
(qw_{2s-1})^{-\lambda_{2s-1}^a-\lambda_{2s}^a-\lambda_s^c}
\prod_{s=1}^{j-1}
(qw_{2s})^{\lambda_{2s}^a+\lambda_{2s+1}^a+\lambda_s^c}
\prod_{s=1}^{2j-1}g_s(w_s)
\frac{\displaystyle \prod_{s=1}^j c_s(q^{\epsilon_{2s-1}}w_{2s-1})}{
\displaystyle
\prod_{s=1}^{j-1}
c_s(w_{2s})}
\nonumber\\
&\times&
\frac{(1-q/zw_1)}{D(z,w_1)}
\prod_{s=1}^{j-1}(1-q/w_{2s}w_{2s+1})
\prod_{s=1}^{j-1}(1-qw_{2s-1}w_{2s})
\nonumber\\
&\times&
\langle \lambda^a;\lambda^c |e^G
e^{Q_{a^{2j}}-Q_{c^j}}e^{B_+(qz;\frac{1}{2})+B_+(q/z;\frac{1}{2})}
\label{action:vo1}\\
&\times&
\prod_{s=1}^{j-1} H_{\epsilon_{2s-1}}^{2s-1}(w_{2s-1},w_{2s})
\prod_{s=1}^{j-1} H^{2s}(w_{2s},w_{2s+1}) J_{\epsilon_{2j-1}}^{2j-1}(w_{2j-1})
~~~~~(j=1,2,\cdots,N),\nonumber
\end{eqnarray}
where we take the integration contour ${C}$ to be simple closed curve that encircles $w_s=0$, $\sqrt{-1}$, $qw_{s-1}^{\pm 1}$
but not $-\sqrt{-1}$, $q^{-1}w_{s-1}^{\pm 1}$ for $s=1,2,\cdots, 2j-1$. 
\begin{eqnarray}
\langle \lambda^a;\lambda^c |e^G \phi_{2j+1}^*(z)
&=&(q-q^{-1})^j
\prod_{s=1}^N e^{\sqrt{-1}\pi \lambda_{2s}^a}
\prod_{s=1}^j\prod_{t=1}^s e^{-\sqrt{-1}\pi \lambda_{2t-1}^a}
\prod_{s=1}^{j}\prod_{t=1}^s e^{-\sqrt{-1}\pi \lambda_{2t}^a}f_1(z)
\nonumber\\
&\times&\sum_{\epsilon_1,\epsilon_3,\cdots,\epsilon_{2j-1}=\pm}
\prod_{s=1}^j \epsilon_{2s-1} 
\prod_{s=1}^{j}q^{-\lambda_s^c \epsilon_{2s-1}}
\prod_{s=1}^{2j} \oint_{{C}} \frac{dw_s}{2\pi \sqrt{-1}w_s}
\nonumber\\
&\times&
\prod_{s=1}^j(qw_{2s-1})^{-\lambda_{2s-1}^a-\lambda_{2s}^a-\lambda_s^c}
\prod_{s=1}^{j}(qw_{2s})^{\lambda_{2s}^a+\lambda_{2s+1}^a+\lambda_s^c}
\prod_{s=1}^{2j}g_s(w_s)
\prod_{s=1}^j \frac{c_s(q^{\epsilon_{2s-1}}w_{2s-1})}{c_s(w_{2s})}
\nonumber\\
&\times&
\frac{(1-q/zw_1)}{D(z,w_1)}
\prod_{s=1}^{j-1}(1-q/w_{2s}w_{2s+1})
\prod_{s=1}^{j}(1-qw_{2s-1}w_{2s})\nonumber\\
&\times&
\langle \lambda^a;\lambda^c |e^G e^{Q_{a^{2j+1}}}e^{B_+(qz;\frac{1}{2})+B_+(q/z;\frac{1}{2})}
\label{action:vo2}\\
&\times&
\prod_{s=1}^j H_{\epsilon_{2s-1}}^{2s-1}(w_{2s-1},w_{2s}) \prod_{s=1}^{j-1}H^{2s}(w_{2s},w_{2s+1})
J^{2j}(w_{2j})
~~~~~(j=0,1,2,\cdots,N-1),\nonumber
\end{eqnarray}
where we take the integration contour ${C}$ to be simple closed curve that encircles $w_s=0$, $\sqrt{-1}$, $qw_{s-1}^{\pm 1}$
but not $-\sqrt{-1}$, $q^{-1}w_{s-1}^{\pm 1}$ for $s=1,2,\cdots, 2j$. 
Here $c_j(w), f_1(z)$, and $g_j(w)$ are given by
(\ref{def:c}), (\ref{def:f}), and (\ref{def:g}), respectively.
\end{prop}
We understand (\ref{action:vo2}) for $j=0$ as follows.
\begin{eqnarray}
\langle \lambda^a;\lambda^c |e^G \phi_1^*(z)=\prod_{s=1}^N e^{\sqrt{-1}\pi \lambda_{2s}^a}
f_1(z)\langle \lambda^a;\lambda^c|e^G e^{Q_{a_1}} e^{B_+(qz;\frac{1}{2})+B_+(q/z;\frac{1}{2})}.\nonumber
\end{eqnarray}

\subsection{Proof for $U_q(\widehat{gl}(1|1))$}

First we show the case $U_q(\widehat{gl}(1|1))$ for simplicity.
It is enough to consider the case 
$
K(z)=\frac{\varphi^{[2]}(z)}{\varphi^{[2]}(z^{-1})}
\left(
\begin{array}{cc}
\frac{1-rz}{1-r/z}&0\\
0&1
\end{array}\right)$.
We would like to show the following two relations coming from the sufficient condition (\ref{eqn:sufficient}).

\begin{prop}
\begin{eqnarray}
(1-rz) \varphi^{[2]}(z)\langle 0|e^G \phi_1^*(z^{-1})
&=&(1-r/z) \varphi^{[2]}(z^{-1}) \langle 0|e^G \phi_1^*(z),
\label{proof:N=1:1}\\
\varphi^{[2]}(z) \langle 0|e^G \phi_2^*(z^{-1})
&=&\varphi^{[2]}(z^{-1}) \langle 0|e^G \phi_2^*(z).
\label{proof:N=1:2}
\end{eqnarray}
\end{prop}
{\it Proof.}~~~\\
$\bullet$~The relation (\ref{proof:N=1:1}) :
From (\ref{action:vo2}), we have
\begin{eqnarray}
&&(1-rz)\varphi^{[2]}(z) \langle 0|e^G \phi_1^*(z^{-1})\nonumber\\
&=&(1-rz)(1-r/z)
\varphi^{[2]}(z)\varphi^{[2]}(z^{-1})
e^{\sqrt{-1}\pi \lambda_2^a} \langle 0|
e^G e^{Q_{a^1}}e^{B_+(qz;\frac{1}{2})+B_+(q/z;\frac{1}{2})}\\
&=&(1-r/z) \varphi^{[2]}(z^{-1}) \langle 0|e^G \phi_1^*(z).\nonumber
\end{eqnarray}
We have shown the first relation (\ref{proof:N=1:1}).
\\
$\bullet$~The relation (\ref{proof:N=1:2}) : 
From (\ref{action:vo1}), we have LHS of (\ref{proof:N=1:2}) as follows.
\begin{eqnarray}
{\rm LHS}&=&-e^{\sqrt{-1}\pi (-\lambda_1^a+\lambda_2^a)}\varphi(z)\varphi(z^{-1})
\langle 0|e^G e^{Q_{a^2}-Q_{c^1}}e^{B_+(qz;\frac{1}{2})+B_+(q/z;\frac{1}{2})}\nonumber\\
&\times&\sum_{\epsilon=\pm}\epsilon \oint \frac{dw}{2\pi \sqrt{-1}w}
\frac{(1-qz/w)(1-r/z)}{D(z,w)}g_1(w)
q^{-\epsilon} c_1(q^{\epsilon}w) J_\epsilon^1(w).
\end{eqnarray}
Because ${\rm RHS}={\rm LHS}|_{z \to 1/z}$, we have
\begin{eqnarray}
{\rm LHS}-{\rm RHS}&=&qe^{\sqrt{-1}\pi (-\lambda_1^a+\lambda_2^a)}(z-z^{-1})\varphi(z)\varphi(z^{-1})
\langle 0|e^G e^{Q_{a^2}-Q_{c^1}}e^{B_+(qz;\frac{1}{2})+B_+(q/z;\frac{1}{2})}\nonumber\\
&\times&\sum_{\epsilon=\pm} \epsilon \oint \frac{dw}{2\pi \sqrt{-1}w}\frac{1}{D(z,w)}
\left(q^\epsilon w+\frac{1}{q^\epsilon w}\right)J_\epsilon^1(w).
\end{eqnarray}
Here we have used $D(z,w)=D(z^{-1},w)$ and $g_1(w)=\frac{1}{1-rw/q}, c_1(w)=1+w^2$.
Taking into account of the relation
$\oint \frac{dw}{w}f(w)=\frac{1}{2}\oint \frac{dw}{w}(f(w)+f(w^{-1}))$, we have
\begin{eqnarray}
\sum_{\epsilon=\pm}\epsilon \oint \frac{dw}{w}\frac{q^\epsilon w+(q^\epsilon w)^{-1}}{D(z,w)}J_\epsilon^1(w)=
\sum_{\epsilon=\pm}\epsilon \oint \frac{dw}{2w}\frac{(q^\epsilon w)+(q^\epsilon w)^{-1}}{D(z,w)}(J_\epsilon^1(w)+J_\epsilon (w^{-1}))
=0.
\nonumber
\end{eqnarray}
Here we have used $D(z,w)=D(z,w^{-1})$ and
$J_\epsilon^1(w^{-1})=J_{-\epsilon}^1(w)$.
Hence we have ${\rm LHS}={\rm RHS}$. We have shown (\ref{proof:N=1:2}).~~~{\bf Q.E.D.}

\subsection{Proof for $U_q(\widehat{gl}(N|N))$}

In this Section we show the case of higher-rank $U_q(\widehat{gl}(N|N))$ $(N \geq 2)$.
Upon the specialization $r=0$
for the case $L=0, M,R>0$ such that $M+R=2N$, we have
${\rm diag}
\left.\left(\frac{1-rz}{1-r/z},\cdots,\frac{1-rz}{1-r/z},
1,\cdots,1\right)\right|_{r=0}=id$.
Hence, it is enough to consider 
two cases $K(z)=\frac{\varphi^{[s]}(z)}{\varphi^{[s]}(z^{-1})}
\overline{K}(z)~~(s=2,3)$. 

\begin{prop}~~~\label{prop:proof1}
For $L=0, M,R>0$ such that $M+R=2N$, we have
\begin{eqnarray}
(1-rz) \varphi^{[2]}(z) \langle 0|e^G \phi_i^*(z^{-1})&=&
(1-r/z) \varphi^{[2]}(z^{-1})\langle 0|e^G \phi_i^*(z)~~(1\leq i \leq M),
\label{proof:N2:1}
\\
\varphi^{[2]}(z)\langle 0|e^G \phi_i^*(z^{-1})&=&
\varphi^{[2]}(z^{-1})\langle 0|e^G \phi_i^*(z)~~~~~~~~~~(M<i \leq 2N).
\label{proof:N2:2}
\end{eqnarray}
\end{prop}

\begin{prop}~~~\label{prop:proof2}
For $L,M,R>0$ such that $L+M+R=2N$, we have
\begin{eqnarray}
z \varphi^{[3]}(z) \langle 0|e^G \phi_i^*(z^{-1})&=&
z^{-1} \varphi^{[3]}(z^{-1})\langle 0|e^G\phi_i^*(z)~~~~~(1\leq i \leq L),
\label{proof:N3:1}
\\
(1-rz)\varphi^{[3]}(z)\langle 0|e^G \phi_i^*(z^{-1})&=&
(1-r/z)\varphi^{[3]}(z^{-1})\langle 0|e^G\phi_i^*(z)~(L< i \leq L+M),
\label{proof:N3:2}
\\
\varphi^{[3]}(z)\langle 0|e^G \phi_i^*(z^{-1})&=&
\varphi^{[3]}(z^{-1})\langle 0|e^G \phi_i^*(z)~~~~~~~~~~(L+M< i \leq 2N).
\label{proof:N3:3}
\end{eqnarray}
\end{prop}
{\it Proof of Proposition \ref{prop:proof1}.}~~~\\
$\bullet$~The relation (\ref{proof:N2:1}) for $i=1$ :
From (\ref{action:vo2}), we have LHS and RHS of (\ref{proof:N2:1}) as follows.
\begin{eqnarray}
{\rm LHS}=(1-rz)(1-r/z)\varphi^{[2]}(z)\varphi^{[2]}(z^{-1})
\langle 0|e^G e^{Q_{a^1}}e^{B_+(qz;\frac{1}{2})+B_+(q/z;\frac{1}{2})}={\rm RHS}.
\end{eqnarray}
We have shown (\ref{proof:N2:1}) for $i=1$.
\\
$\bullet$~The relation (\ref{proof:N2:1}) for $i=2j$ $(j \geq 1)$ :
From (\ref{action:vo1}), we have ${\rm LHS}-{\rm RHS}$ of (\ref{proof:N2:1}) for $i=2j$ as follows.
\begin{eqnarray}
&&{\rm LHS}-{\rm RHS}\nonumber\\
&=&q(q-q^{-1})^{j-1}
(z-z^{-1})(1-rz)(1-r/z) \varphi^{[2]}(z)\varphi^{[2]}(z^{-1})\nonumber\\
&\times&
\sum_{\epsilon_1,\cdots,\epsilon_{2j-1}=\pm} \prod_{s=1}^j \epsilon_{2s-1} q^{-\epsilon_{2j-1}}
\prod_{s=1}^{2j-1}\oint_C \frac{dw_s}{2\pi \sqrt{-1}w_s} \prod_{s=1}^{2j-1} g_s(w_s)
\frac{\prod_{s=1}^j c_s(q^{\epsilon_{2s-1}}w_{2s-1})}{\prod_{s=1}^{j-1}c_s(w_{2s})}\nonumber\\
&\times&
\frac{\prod_{s=1}^{j-1}(1-q/w_{2s}w_{2s+1})(1-qw_{2s-1}w_{2s})}{w_1 D(z,w_1)}
\langle 0|e^Ge^{Q_{a^{2j}-c^j}} e^{B_+(qz;\frac{1}{2})+B_+(q/z;\frac{1}{2})}\nonumber\\
&\times& \prod_{s=1}^{j-1}H_{\epsilon_{2s-1}}^{2s-1}(w_{2s-1},w_{2s})
\prod_{s=1}^{j-1} H^{2s}(w_{2s},w_{2s+1}) J_{\epsilon_{2j-1}}^{2j-1}(w_{2j-1}).
\label{induction:0}
\end{eqnarray}
We focus our attention to the following integral relating to the variables $w_1,w_2,\cdots,w_{2j-1}$
in (\ref{induction:0}). 
\begin{eqnarray}
&&\sum_{\epsilon_1,\cdots,\epsilon_{2j-1}=\pm} \prod_{s=1}^j \epsilon_{2s-1} q^{-\epsilon_{2j-1}}
\prod_{s=1}^{2j-1}\oint_C \frac{dw_s}{2\pi \sqrt{-1}w_s} \prod_{s=1}^{2j-1} g_s(w_s)
\frac{\prod_{s=1}^j c_s(q^{\epsilon_{2s-1}}w_{2s-1})}{\prod_{s=1}^{j-1}c_s(w_{2s})}
\nonumber
\\
&\times&
\prod_{s=1}^{j-1}(1-q/w_{2s}w_{2s+1})(1-qw_{2s-1}w_{2s})
\prod_{s=1}^{j-1}H_{\epsilon_{2s-1}}^{2s-1}(w_{2s-1},w_{2s})
\prod_{s=1}^{j-1}H^{2s}(w_{2s},w_{2s+1})\nonumber\\
&\times& J_{\epsilon_{2j-1}}^{2j-1}(w_{2j-1})
\frac{1}{w_1 D(z,w_1)}.
\label{induction:1}
\end{eqnarray}

First we study (\ref{induction:1}) for $j=1$.
We have
\begin{eqnarray}
\sum_{\epsilon_1=\pm}\epsilon_1 \oint \frac{dw_1}{w_1}\frac{q^{-\epsilon_1}c_1(q^{\epsilon_1}w_1)}{w_1}
\frac{1}{D(z,w_1)}J_{\epsilon_1}^1(w_1)=0.\nonumber
\end{eqnarray}
Now we have shown (\ref{induction:1}) for $j=1$.

Next we study (\ref{induction:1}) for $j \geq 2$. 
We focus our attention to the variable $w_1$.
Using (\ref{eqn:recursion1}) and $D(z,w_1)=D(z,w_1^{-1})$, we have
\begin{eqnarray}
&&
\sum_{\epsilon_1=\pm}\epsilon_1 \oint \frac{dw_1}{w_1} \frac{c_1(q^{\epsilon_1}w_1)}{w_1} \frac{(1-qw_1w_2)}{D(z,w_1)}H_{\epsilon_1}^1(w_1,w_2)\nonumber\\
&=&
\oint \frac{dw_1}{w_1} q\left(qw_1+\frac{1}{qw_1}\right)\frac{(1-w_1^2)}{D(z,w_1)}H_+^1(w_1,w_2).
\nonumber
\end{eqnarray}
After calculation for $w_1$ we focus our attention to the variable $w_2$. 
Using (\ref{eqn:recursion5}) and $H_+^1(w_1^{-1},w_2)=H_+^1(w_1,w_2)$ we have
\begin{eqnarray}
&&
\oint \frac{dw_2}{w_2}(1-q/w_2w_3)\frac{g_2(w_2)}{c_2(w_2)}H^2(w_2,w_3)H_+^1(w_1,w_2)\nonumber\\
&=&
-\frac{q}{2w_3} \oint \frac{dw_2}{w_2}(1-q/w_2w_3)\frac{(1-w_2^2)^2}{w_2(1+w_2^2)}H^2(w_2,w_3)H_+^1(w_1,w_2).
\nonumber
\end{eqnarray}
We get the following integral relating to
the variables $w_3, w_4, \cdots, w_{2j-1}$. 
\begin{eqnarray}
&&\sum_{\epsilon_3,\cdots,\epsilon_{2j-1}=\pm} 
\prod_{s=2}^j \epsilon_{2s-1} q^{-\epsilon_{2j-1}}
\prod_{s=3}^{2j-1}\oint_C \frac{dw_s}{2\pi \sqrt{-1}w_s} \prod_{s=3}^{2j-1} g_s(w_s)
\frac{\prod_{s=2}^j c_s(q^{\epsilon_{2s-1}}w_{2s-1})}{\prod_{s=2}^{j-1}c_s(w_{2s})}
\nonumber
\\
&\times&
\prod_{s=2}^{j-1}(1-q/w_{2s}w_{2s+1})(1-qw_{2s-1}w_{2s})
\prod_{s=2}^{j-1}H_{\epsilon_{2s-1}}^{2s-1}(w_{2s-1},w_{2s})H^{2s}(w_{2s},w_{2s+1})\nonumber\\
&\times& J_{\epsilon_{2j-1}}^{2j-1}(w_{2j-1}) \frac{1}{w_3}H^2(w_2,w_3).
\label{induction:2}
\end{eqnarray}
The structure of (\ref{induction:1}) and (\ref{induction:2}) are the same except for
their sizes and minor difference between
$D(z,w_1)$ and $H^2(w_2,w_3)$. 
We note that $D(z,w)$ and $H^2(z,w)$ are invariant under $w \to w^{-1}$. 
Using (\ref{eqn:recursion1}) and (\ref{eqn:recursion6}) 
we calculate the variables $w_3,w_4,\cdots,w_{2j-2}$ iteratively.
Then we get the same relation as (\ref{induction:1}) for $j=1$,
that we have already shown.
\\
$\bullet$~
The relation (\ref{proof:N2:1}) for
$i=2j+1$ $(j \geq 1)$ :
From (\ref{action:vo2}), we have ${\rm LHS}-{\rm RHS}$ of (\ref{proof:N2:1}) for $i=2j+1$ as follows.
\begin{eqnarray}
&&{\rm LHS}-{\rm RHS}\nonumber\\
&=&-q(q-q^{-1})^{j}
(z-z^{-1})(1-rz)(1-r/z) \varphi^{[2]}(z)\varphi^{[2]}(z^{-1})\nonumber\\
&\times&
\sum_{\epsilon_1,\cdots,\epsilon_{2j-1}=\pm} \prod_{s=1}^j \epsilon_{2s-1}
\prod_{s=1}^{2j}
\oint_C \frac{dw_s}{2\pi \sqrt{-1}w_s} 
\prod_{s=1}^{2j} g_s(w_s)
\prod_{s=1}^j
\frac{c_s(q^{\epsilon_{2s-1}}w_{2s-1})}{c_s(w_{2s})}\nonumber\\
&\times&
\frac{\prod_{s=1}^{j-1}(1-q/w_{2s}w_{2s+1})\prod_{s=1}^j (1-qw_{2s-1}w_{2s})}{w_1 D(z,w_1)}
\langle 0|e^Ge^{Q_{a^{2j}}} e^{B_+(qz;\frac{1}{2})+B_+(q/z;\frac{1}{2})}\nonumber\\
&\times& \prod_{s=1}^{j}H_{\epsilon_{2s-1}}^{2s-1}(w_{2s-1},w_{2s})
\prod_{s=1}^{j-1} H^{2s}(w_{2s},w_{2s+1}) J^{2j}(w_{2j}).
\label{induction:2.5}
\end{eqnarray}
We focus our attention to the following integral relating to the variables $w_1,w_2,\cdots,w_{2j-1}$
in (\ref{induction:2.5}). 
\begin{eqnarray}
&&\sum_{\epsilon_1,\cdots,\epsilon_{2j-1}=\pm} \prod_{s=1}^j \epsilon_{2s-1}
\prod_{s=1}^{2j}\oint_C \frac{dw_s}{2\pi \sqrt{-1}w_s} 
\prod_{s=1}^{2j} g_s(w_s)
\prod_{s=1}^j \frac{c_s(q^{\epsilon_{2s-1}}w_{2s-1})}{c_s(w_{2s})}
\nonumber
\\
&\times&
\prod_{s=1}^{j-1}(1-q/w_{2s}w_{2s+1})
\prod_{s=1}^j (1-qw_{2s-1}w_{2s})
\prod_{s=1}^{j}H_{\epsilon_{2s-1}}^{2s-1}(w_{2s-1},w_{2s})
\prod_{s=1}^{j-1}H^{2s}(w_{2s},w_{2s+1})\nonumber\\
&\times& J^{2j}(w_{2j})
\frac{1}{w_1 D(z,w_1)}.
\label{induction:3}
\end{eqnarray}

First we study (\ref{induction:3}) for $j=1$. 
We calculate the variable $w_1$ using (\ref{eqn:recursion1}). 
Then we have the following integral relating to $w_2$.
\begin{eqnarray}
\oint \frac{dw_2}{w_2} \frac{g_2(w_2)}{c_2(w_2)}J^2(w_2)=0,\nonumber
\end{eqnarray}
where we have used $J^2(w_2^{-1})=J^2(w_2)$.
Now we have shown 
(\ref{induction:3}) for $j=1$.

Next we study (\ref{induction:3}) for $j \geq 2$.
We focus our attention to two variables $w_1, w_2$.
Using relations (\ref{eqn:recursion1}) and (\ref{eqn:recursion6}) in the same way as the above,
we have the following integral relating to $w_3,w_4,\cdots,w_{2j}$.
\begin{eqnarray}
&&\sum_{\epsilon_3,\cdots,\epsilon_{2j-1}=\pm} \prod_{s=2}^j \epsilon_{2s-1}
\prod_{s=3}^{2j}\oint_C \frac{dw_s}{2\pi \sqrt{-1}w_s} 
\prod_{s=3}^{2j} g_s(w_s)
\prod_{s=2}^j \frac{c_s(q^{\epsilon_{2s-1}}w_{2s-1})}{c_s(w_{2s})}
\nonumber
\\
&\times&
\prod_{s=2}^{j-1}(1-q/w_{2s}w_{2s+1})
\prod_{s=2}^j (1-qw_{2s-1}w_{2s})
\prod_{s=2}^{j}H_{\epsilon_{2s-1}}^{2s-1}(w_{2s-1},w_{2s})
\prod_{s=2}^{j-1}H^{2s}(w_{2s},w_{2s+1})\nonumber\\
&\times& J^{2j}(w_{2j})
\frac{1}{w_3}H^2(w_2,w_3).
\label{induction:4}
\end{eqnarray}
The structure of (\ref{induction:3}) and (\ref{induction:4}) are the same except for
their sizes and minor difference between
$D(z,w_1)$ and $H^2(w_2,w_3)$.
Using (\ref{eqn:recursion1}) and (\ref{eqn:recursion6}) 
we calculate the variables $w_3,w_4,\cdots,w_{2j-2}$ iteratively.
Then we get the same relation as (\ref{induction:3}) for $j=1$,
that we have already shown.
\\
$\bullet$~The relation (\ref{proof:N2:2}) for $i=2j$ :
From (\ref{action:vo1}), we have ${\rm LHS}-{\rm RHS}$ of (\ref{proof:N2:2}) for $i=2j$ as follows.
\begin{eqnarray}
&&{\rm LHS}-{\rm RHS}\nonumber\\
&=&q(q-q^{-1})^{j-1}
(z-z^{-1})\varphi^{[2]}(z)\varphi^{[2]}(z^{-1})\nonumber\\
&\times&
\sum_{\epsilon_1,\cdots,\epsilon_{2j-1}=\pm} \prod_{s=1}^j \epsilon_{2s-1} q^{-\epsilon_{2j-1}}
\prod_{s=1}^{2j-1}\oint_C \frac{dw_s}{2\pi \sqrt{-1}w_s} \prod_{s=1}^{2j-1} g_s(w_s)
\frac{\prod_{s=1}^j c_s(q^{\epsilon_{2s-1}}w_{2s-1})}{\prod_{s=1}^{j-1}c_s(w_{2s})}\nonumber\\
&\times&
(1-rw_1/q) \frac{\prod_{s=1}^{j-1}(1-q/w_{2s}w_{2s+1})(1-qw_{2s-1}w_{2s})}{w_1 D(z,w_1)}
\langle 0|e^Ge^{Q_{a^{2j}-c^j}} e^{B_+(qz;\frac{1}{2})+B_+(q/z;\frac{1}{2})}\nonumber\\
&\times& \prod_{s=1}^{j-1}H_{\epsilon_{2s-1}}^{2s-1}(w_{2s-1},w_{2s})
\prod_{s=1}^{j-1} H^{2s}(w_{2s},w_{2s+1}) J_{\epsilon_{2j-1}}^{2j-1}(w_{2j-1}).
\label{induction:4.5}
\end{eqnarray}
We focus our attention to the following part relating to the variables $w_1,w_2,\cdots,w_{2j-1}$
in (\ref{induction:4.5}). 
\begin{eqnarray}
&&\sum_{\epsilon_1,\cdots,\epsilon_{2j-1}=\pm} \prod_{s=1}^j \epsilon_{2s-1} q^{-\epsilon_{2j-1}}
\prod_{s=1}^{2j-1}\oint_C \frac{dw_s}{2\pi \sqrt{-1}w_s} \prod_{s=1}^{2j-1} g_s(w_s)
\frac{\prod_{s=1}^j c_s(q^{\epsilon_{2s-1}}w_{2s-1})}{\prod_{s=1}^{j-1}c_s(w_{2s})}
\nonumber
\\
&\times&
\prod_{s=1}^{j-1}(1-q/w_{2s}w_{2s+1})(1-qw_{2s-1}w_{2s})
\prod_{s=1}^{j-1}H_{\epsilon_{2s-1}}^{2s-1}(w_{2s-1},w_{2s})
\prod_{s=1}^{j-1}H^{2s}(w_{2s},w_{2s+1})\nonumber\\
&\times& (1-rw_1/q)\times J_{\epsilon_{2j-1}}^{2j-1}(w_{2j-1})
\frac{1}{w_1 D(z,w_1)}.
\label{induction:5}
\end{eqnarray}

First we study (\ref{induction:5}) for $M=1$ and $j=1$, where $g_1(w)=\frac{1}{1-rw/q}$.
\begin{eqnarray}
\sum_{\epsilon_1=\pm} \epsilon_1 q^{-\epsilon_1} \oint \frac{dw_1}{w_1}
\frac{c_1(q^{\epsilon_1}w_1)}{w_1}\frac{(1-rw_1/q)g_1(w_1)}{D(z,w_1)}J_{\epsilon_1}^1(w_1)=0.\nonumber
\end{eqnarray}
We have shown (\ref{induction:5}) for $M=1$ and $j=1$.

Next we study (\ref{induction:5}) for $M=1$ and $j \geq 2$, where $g_1(w)=\frac{1}{1-rw/q}$.
We focus our attention to the variable $w_1$. Using (\ref{eqn:recursion1}), we have
\begin{eqnarray}
&&\sum_{\epsilon_1=\pm}\epsilon_1 \oint \frac{dw_1}{w_1} \frac{c_1(q^{\epsilon_1}w_1)}{w_1}
\frac{g_1(w_1)(1-rw_1/q)}{D(z,w_1)} (1-qw_1w_2) H_{\epsilon_1}^1(w_1,w_2)\nonumber\\
&=&\oint \frac{dw_1}{w_1}q\left(qw_1+\frac{1}{qw_1}\right)\frac{(1-w_1^2)}{D(z,w_1)}H_+^1(w_1,w_2).\nonumber
\end{eqnarray}
We focus our attention to the variable $w_2$.
Using (\ref{eqn:recursion6}) we have (\ref{induction:2}) as the part relating to $w_3,w_4,\cdots,w_{2j}$.
Hence we have shown (\ref{induction:5}) for $M=1$ and $j \geq 2$.

Finally we study the case $M \geq 2$ and $j \geq 2$.
Using (\ref{eqn:recursion2}), (\ref{eqn:recursion6})
we calculate $w_1,w_2,\cdots,w_{M-1}$ iteratively.
Hence we have the following integrals relating to $w_M,w_{M+1},\cdots,w_{2j-1}$.
For $M=$ odd, we have
\begin{eqnarray}
&&\sum_{\epsilon_M,\cdots,\epsilon_{2j-1}=\pm} 
\prod_{s=\frac{M+1}{2}}^j \epsilon_{2s-1} q^{-\epsilon_{2j-1}}
\prod_{s=M}^{2j-1}\oint_C \frac{dw_s}{2\pi \sqrt{-1}w_s} 
\prod_{s=M}^{2j-1} g_s(w_s)
\frac{\prod_{s=\frac{M+1}{2}}^j c_s(q^{\epsilon_{2s-1}}w_{2s-1})}
{\prod_{s=\frac{M+1}{2}}^{j-1}c_s(w_{2s})}
\nonumber
\\
&\times&
\prod_{s=\frac{M+1}{2}}^{j-1}(1-q/w_{2s}w_{2s+1})(1-qw_{2s-1}w_{2s})
\prod_{s=\frac{M+1}{2}}^{j-1}H_{\epsilon_{2s-1}}^{2s-1}(w_{2s-1},w_{2s})
\prod_{s=\frac{M+1}{2}}^{j-1}H^{2s}(w_{2s},w_{2s+1})\nonumber\\
&\times&(1-rw_M/q) \times J_{\epsilon_{2j-1}}^{2j-1}(w_{2j-1})
\frac{H^{M-1}(w_{M-1},w_M)}{w_M}.
\label{induction:6}
\end{eqnarray}
For $M=$ even, we have
\begin{eqnarray}
&&\sum_{\epsilon_{M+1},\cdots,\epsilon_{2j-1}=\pm} 
\prod_{s=\frac{M}{2}+1}^j \epsilon_{2s-1} q^{-\epsilon_{2j-1}}
\prod_{s=M}^{2j-1}\oint_C \frac{dw_s}{2\pi \sqrt{-1}w_s} 
\prod_{s=M}^{2j-1} g_s(w_s)
\frac{\prod_{s=\frac{M}{2}+1}^j c_s(q^{\epsilon_{2s-1}}w_{2s-1})}{\prod_{s=\frac{M}{2}}^{j-1}c_s(w_{2s})}
\nonumber
\\
&\times&
\prod_{s=\frac{M}{2}}^{j-1}(1-q/w_{2s}w_{2s+1})
\prod_{s=\frac{M}{2}+1}^{j-1}(1-qw_{2s-1}w_{2s})
\prod_{s=\frac{M}{2}+1}^{j-1}H_{\epsilon_{2s-1}}^{2s-1}(w_{2s-1},w_{2s})
\prod_{s=\frac{M}{2}}^{j-1}H^{2s}(w_{2s},w_{2s+1})\nonumber\\
&\times& (1-rw_M)\times J_{\epsilon_{2j-1}}^{2j-1}(w_{2j-1})
\frac{H_{+}^{M-1}(w_{M-1},w_M)}{w_M}.
\label{induction:7}
\end{eqnarray}
For $M=$ odd, we have $g_M(w)=\frac{1}{(1-rw/q)}$.
Hence (\ref{induction:6}) becomes the same as (\ref{induction:2})
except for its size. Hence we have shown the case for $M\geq 2$, $j \geq2$, and $M=$ odd. 
For $M=$ even, we have $g_M(w)=\frac{(1-w^2)}{(1-rw)}$.
Using (\ref{eqn:recursion5}) we calculate the variable $w_M$.
Hence (\ref{induction:7}) becomes the same as (\ref{induction:2})
except for its size. Hence we have shown the case for $M\geq 2$, $j \geq2$, and $M=$ even. 
\\
$\bullet$~The relation (\ref{proof:N2:2}) for $i=2j+1$ :
From (\ref{action:vo2}), we have ${\rm LHS}-{\rm RHS}$ of (\ref{proof:N2:2}) for $i=2j+1$ as follows.
\begin{eqnarray}
&&{\rm LHS}-{\rm RHS}\nonumber\\
&=&-q(q-q^{-1})^{j}
(z-z^{-1}) \varphi^{[2]}(z)\varphi^{[2]}(z^{-1})\nonumber\\
&\times&
\sum_{\epsilon_1,\cdots,\epsilon_{2j-1}=\pm} \prod_{s=1}^j \epsilon_{2s-1}
\prod_{s=1}^{2j}
\oint_C \frac{dw_s}{2\pi \sqrt{-1}w_s} 
\prod_{s=1}^{2j} g_s(w_s)
\prod_{s=1}^j
\frac{c_s(q^{\epsilon_{2s-1}}w_{2s-1})}{c_s(w_{2s})}(1-rw_1/q)\nonumber\\
&\times&
\frac{\prod_{s=1}^{j-1}(1-q/w_{2s}w_{2s+1})\prod_{s=1}^j (1-qw_{2s-1}w_{2s})}{w_1 D(z,w_1)}
\langle 0|e^Ge^{Q_{a^{2j}}} e^{B_+(qz;\frac{1}{2})+B_+(q/z;\frac{1}{2})}\nonumber\\
&\times& \prod_{s=1}^{j}H_{\epsilon_{2s-1}}^{2s-1}(w_{2s-1},w_{2s})
\prod_{s=1}^{j-1} H^{2s}(w_{2s},w_{2s+1}) J^{2j}(w_{2j}).\label{induction:7.5}
\end{eqnarray}
We focus our attention to the following integral relating to the variables $w_1,w_2,\cdots,w_{2j-1}$
in (\ref{induction:7.5}). 
\begin{eqnarray}
&&\sum_{\epsilon_1,\cdots,\epsilon_{2j-1}=\pm} \prod_{s=1}^j \epsilon_{2s-1}
\prod_{s=1}^{2j}\oint_C \frac{dw_s}{2\pi \sqrt{-1}w_s} 
\prod_{s=1}^{2j} g_s(w_s)
\prod_{s=1}^j \frac{c_s(q^{\epsilon_{2s-1}}w_{2s-1})}{c_s(w_{2s})}
\nonumber
\\
&\times&
\prod_{s=1}^{j-1}(1-q/w_{2s}w_{2s+1})
\prod_{s=1}^j (1-qw_{2s-1}w_{2s})
\prod_{s=1}^{j}H_{\epsilon_{2s-1}}^{2s-1}(w_{2s-1},w_{2s})
\prod_{s=1}^{j-1}H^{2s}(w_{2s},w_{2s+1})\nonumber\\
&\times& (1-rw_1/q)\times J^{2j}(w_{2j})
\frac{1}{w_1 D(z,w_1)}.
\label{induction:8}
\end{eqnarray}

First we study (\ref{induction:8}) for $M=1$ and $j \geq 1$,
where $g_1(w)=\frac{1}{(1-rw/q)}$. 
We focus our attention to the variable $w_1$.
Using (\ref{eqn:recursion2}) we have
\begin{eqnarray}
&&\sum_{\epsilon_1=\pm}\epsilon_1\oint \frac{dw_1}{w_1}
\frac{c_1(q^{\epsilon_1}w_1)}{w_1}\frac{(1-qw_1w_2)}{D(z,w_1)}g_1(w_1)(1-rw_1/q)H_{\epsilon_1}^1(w_1,w_2)\nonumber\\
&=&\oint \frac{dw_1}{w_1} q\left(qw_1+\frac{1}{qw_1}\right)\frac{(1-w_1^2)}{D(z,w_1)}H_+^1(w_1,w_2).\nonumber
\end{eqnarray}
We focus our attention to the variable $w_2$.
Using (\ref{eqn:recursion6}) we have
\begin{eqnarray}
&&\oint \frac{dw_2}{w_2} \frac{g_2(w_2)}{c_2(w_2)}(1-q/w_2w_3)H^2(w_2,w_3)H_+^1(w_1,w_2)\nonumber\\
&=&
-\frac{q}{2w_3}\oint \frac{dw_2}{w_2}
\frac{(1-w_2^2)^2}{w_2(1+w_2^2)}H^2(w_2,w_3)H_+^1(w_1,w_2).\nonumber
\end{eqnarray}
Then we have (\ref{induction:4}) as the part relation to $w_3,w_4,\cdots,w_{2j-1}$.
Hence we have shown (\ref{induction:8}) for $M=1$ and $j \geq 1$.

Next we study (\ref{induction:8}) for $M \geq 2$ and $j \geq 1$. 
Using (\ref{eqn:recursion2}) and (\ref{eqn:recursion5}) we have
the following integrals relating to the variables $w_M, w_{M+1},\cdots, w_{2j}$.
For $M=$ odd, we have
\begin{eqnarray}
&&\sum_{\epsilon_M,\cdots,\epsilon_{2j-1}=\pm} \prod_{s=\frac{M+1}{2}}^j \epsilon_{2s-1}
\prod_{s=M}^{2j}\oint_C \frac{dw_s}{2\pi \sqrt{-1}w_s} 
\prod_{s=M}^{2j} g_s(w_s)
\prod_{s=\frac{M+2}{2}}^j 
\frac{c_s(q^{\epsilon_{2s-1}}w_{2s-1})}{c_s(w_{2s})}
\nonumber
\\
&\times&
\prod_{s=\frac{M+1}{2}}^{j-1}(1-q/w_{2s}w_{2s+1})
\prod_{s=\frac{M+1}{2}}^j (1-qw_{2s-1}w_{2s})
\prod_{s=\frac{M+1}{2}}^{j}H_{\epsilon_{2s-1}}^{2s-1}(w_{2s-1},w_{2s})
\prod_{s=\frac{M+2}{2}}^{j-1}H^{2s}(w_{2s},w_{2s+1})\nonumber\\
&\times& (1-rw_M/q)\times J^{2j}(w_{2j})
\frac{H^{M-1}(w_{M-1},w_M)}{w_M }.
\label{induction:9}
\end{eqnarray}
For $M=$ even, we have
\begin{eqnarray}
&&\sum_{\epsilon_{M+1},\cdots,\epsilon_{2j-1}=\pm} 
\prod_{s=\frac{M}{2}+1}^j \epsilon_{2s-1}
\prod_{s=M}^{2j}\oint_C \frac{dw_s}{2\pi \sqrt{-1}w_s} 
\prod_{s=M}^{2j} g_s(w_s)
\frac{ \prod_{s=\frac{M}{2}+1}^j c_s(q^{\epsilon_{2s-1}}w_{2s-1})}{
\prod_{s=\frac{M}{2}}^jc_s(w_{2s})}
\nonumber
\\
&\times&
\prod_{s=\frac{M}{2}}^{j-1}(1-q/w_{2s}w_{2s+1})
\prod_{s=\frac{M}{2}+1}^j (1-qw_{2s-1}w_{2s})
\prod_{s=\frac{M}{2}+1}^{j}H_{\epsilon_{2s-1}}^{2s-1}(w_{2s-1},w_{2s})
\prod_{s=\frac{M}{2}}^{j-1}H^{2s}(w_{2s},w_{2s+1})\nonumber\\
&\times& (1-rw_M)\times J^{2j}(w_{2j})
\frac{H_+^{M-1}(w_{M-1},w_M)}{w_M}.
\label{induction:10}
\end{eqnarray}
For $M=$ odd, we have $g_M(w)=\frac{1}{(1-rw/q)}$.
Hence (\ref{induction:9}) becomes the same as (\ref{induction:4})
except for its size. Hence we have shown (\ref{induction:9}) for $M\geq 2$, $j \geq 1$, and $M=$ odd. 
For $M=$ even, we have $g_M(w)=\frac{(1-w^2)}{(1-rw)}$.
Using (\ref{eqn:recursion5}) we calculate the variable $w_M$.
Hence (\ref{induction:10}) becomes the same as (\ref{induction:4})
except for its size. Hence we have shown (\ref{induction:10}) for $M\geq 2$, $j \geq 1$, and $M=$ even. ~~~
{\bf Q.E.D.}

~\\
{\it Proof of Proposition \ref{prop:proof2}.}~~~\\
Proposition \ref{prop:proof2} is shown in the same way as Proposition \ref{prop:proof1}.\\
$\bullet$~The relation (\ref{proof:N3:1}) for $j=2i$ :
From (\ref{action:vo1}) the relation (\ref{proof:N3:1}) is reduced to the following relation.
\begin{eqnarray}
&&\sum_{\epsilon_1,\cdots,\epsilon_{2j-1}=\pm} \prod_{s=1}^j \epsilon_{2s-1} q^{-\epsilon_{2j-1}}
\prod_{s=1}^{2j-1}\oint_C \frac{dw_s}{2\pi \sqrt{-1}w_s} \prod_{s=1}^{2j-1} g_s(w_s)
\frac{\prod_{s=1}^j c_s(q^{\epsilon_{2s-1}}w_{2s-1})}{\prod_{s=1}^{j-1}c_s(w_{2s})}
\nonumber
\\
&\times&
\prod_{s=1}^{j-1}(1-q/w_{2s}w_{2s+1})(1-qw_{2s-1}w_{2s})
\prod_{s=1}^{j-1}H_{\epsilon_{2s-1}}^{2s-1}(w_{2s-1},w_{2s})
\prod_{s=1}^{j-1}H^{2s}(w_{2s},w_{2s+1})\nonumber\\
&\times& J_{\epsilon_{2j-1}}^{2j-1}(w_{2j-1})
\frac{1}{w_1 D(z,w_1)}=0.
\label{induction:11}
\end{eqnarray}
Using (\ref{eqn:recursion1}) and (\ref{eqn:recursion6}),
the relation (\ref{induction:11}) is shown in the same way as Proposition \ref{prop:proof1}.
\\
$\bullet$~The relation (\ref{proof:N3:1}) for $j=2i+1$ :
From (\ref{action:vo2}) the relation (\ref{proof:N3:1}) is reduced to the following relation.
\begin{eqnarray}
&&\sum_{\epsilon_1,\cdots,\epsilon_{2j-1}=\pm} \prod_{s=1}^j \epsilon_{2s-1}
\prod_{s=1}^{2j}\oint_C \frac{dw_s}{2\pi \sqrt{-1}w_s} 
\prod_{s=1}^{2j} g_s(w_s)
\prod_{s=1}^j \frac{c_s(q^{\epsilon_{2s-1}}w_{2s-1})}{c_s(w_{2s})}
\nonumber
\\
&\times&
\prod_{s=1}^{j-1}(1-q/w_{2s}w_{2s+1})
\prod_{s=1}^j (1-qw_{2s-1}w_{2s})
\prod_{s=1}^{j}H_{\epsilon_{2s-1}}^{2s-1}(w_{2s-1},w_{2s})
\prod_{s=1}^{j-1}H^{2s}(w_{2s},w_{2s+1})\nonumber\\
&\times& J^{2j}(w_{2j})
\frac{1}{w_1 D(z,w_1)}=0.
\label{induction:12}
\end{eqnarray}
Using (\ref{eqn:recursion1}) and (\ref{eqn:recursion6}),
the relation (\ref{induction:12}) is shown in the same way as Proposition \ref{prop:proof1}.
\\
$\bullet$~The relation (\ref{proof:N3:2}) for $j=2i$ :
From (\ref{action:vo1}) the relation (\ref{proof:N3:2}) is reduced to the following relation.
\begin{eqnarray}
&&\sum_{\epsilon_1,\cdots,\epsilon_{2j-1}=\pm} \prod_{s=1}^j \epsilon_{2s-1} q^{-\epsilon_{2j-1}}
\prod_{s=1}^{2j-1}\oint_C \frac{dw_s}{2\pi \sqrt{-1}w_s} \prod_{s=1}^{2j-1} g_s(w_s)
\frac{\prod_{s=1}^j c_s(q^{\epsilon_{2s-1}}w_{2s-1})}{\prod_{s=1}^{j-1}c_s(w_{2s})}
\nonumber
\\
&\times&
\prod_{s=1}^{j-1}(1-q/w_{2s}w_{2s+1})(1-qw_{2s-1}w_{2s})
\prod_{s=1}^{j-1}H_{\epsilon_{2s-1}}^{2s-1}(w_{2s-1},w_{2s})
\prod_{s=1}^{j-1}H^{2s}(w_{2s},w_{2s+1})\nonumber\\
&\times& (1-w_1/qr) \times J_{\epsilon_{2j-1}}^{2j-1}(w_{2j-1})
\frac{1}{w_1 D(z,w_1)}=0.
\label{induction:13}
\end{eqnarray}
Using (\ref{eqn:recursion1}),  (\ref{eqn:recursion2}), (\ref{eqn:recursion5}), (\ref{eqn:recursion6}),
the relation (\ref{induction:13}) is shown in the same way as Proposition \ref{prop:proof1}.
\\
$\bullet$~The relation (\ref{proof:N3:2}) for $j=2i+1$ :
From (\ref{action:vo2}) the relation (\ref{proof:N3:2}) is reduced to the following relation.
\begin{eqnarray}
&&\sum_{\epsilon_1,\cdots,\epsilon_{2j-1}=\pm} \prod_{s=1}^j \epsilon_{2s-1}
\prod_{s=1}^{2j}\oint_C \frac{dw_s}{2\pi \sqrt{-1}w_s} 
\prod_{s=1}^{2j} g_s(w_s)
\prod_{s=1}^j \frac{c_s(q^{\epsilon_{2s-1}}w_{2s-1})}{c_s(w_{2s})}
\nonumber
\\
&\times&
\prod_{s=1}^{j-1}(1-q/w_{2s}w_{2s+1})
\prod_{s=1}^j (1-qw_{2s-1}w_{2s})
\prod_{s=1}^{j}H_{\epsilon_{2s-1}}^{2s-1}(w_{2s-1},w_{2s})
\prod_{s=1}^{j-1}H^{2s}(w_{2s},w_{2s+1})\nonumber\\
&\times& (1-w_1/qr) \times J^{2j}(w_{2j})
\frac{1}{w_1 D(z,w_1)}=0.
\label{induction:14}
\end{eqnarray}
Using (\ref{eqn:recursion1}),  (\ref{eqn:recursion2}), (\ref{eqn:recursion5}), (\ref{eqn:recursion6}),
the relation (\ref{induction:14}) is shown in the same way as Proposition \ref{prop:proof1}.
\\
$\bullet$~The relation (\ref{proof:N3:3}) for $j=2i$ :
From (\ref{action:vo1}) the relation (\ref{proof:N3:3}) is reduced to the following relation.
\begin{eqnarray}
&&\sum_{\epsilon_1,\cdots,\epsilon_{2j-1}=\pm} \prod_{s=1}^j \epsilon_{2s-1} q^{-\epsilon_{2j-1}}
\prod_{s=1}^{2j-1}\oint_C \frac{dw_s}{2\pi \sqrt{-1}w_s} \prod_{s=1}^{2j-1} g_s(w_s)
\frac{\prod_{s=1}^j c_s(q^{\epsilon_{2s-1}}w_{2s-1})}{\prod_{s=1}^{j-1}c_s(w_{2s})}
\nonumber
\\
&\times&
\prod_{s=1}^{j-1}(1-q/w_{2s}w_{2s+1})(1-qw_{2s-1}w_{2s})
\prod_{s=1}^{j-1}H_{\epsilon_{2s-1}}^{2s-1}(w_{2s-1},w_{2s})
\prod_{s=1}^{j-1}H^{2s}(w_{2s},w_{2s+1})\nonumber\\
&\times& J_{\epsilon_{2j-1}}^{2j-1}(w_{2j-1})
\frac{1}{D(z,w_1)}=0.
\label{induction:15}
\end{eqnarray}
Using 
(\ref{eqn:recursion1}),
(\ref{eqn:recursion2}),  (\ref{eqn:recursion3}), (\ref{eqn:recursion4}), (\ref{eqn:recursion5}), (\ref{eqn:recursion6}),
(\ref{eqn:recursion7}),
the relation (\ref{induction:15}) is shown in the same way as Proposition \ref{prop:proof1}.
\\
$\bullet$~The relation (\ref{proof:N3:3}) for $j=2i+1$ :
From (\ref{action:vo2}) the relation (\ref{proof:N3:3}) is reduced to the following relation.
\begin{eqnarray}
&&\sum_{\epsilon_1,\cdots,\epsilon_{2j-1}=\pm} \prod_{s=1}^j \epsilon_{2s-1}
\prod_{s=1}^{2j}\oint_C \frac{dw_s}{2\pi \sqrt{-1}w_s} 
\prod_{s=1}^{2j} g_s(w_s)
\prod_{s=1}^j \frac{c_s(q^{\epsilon_{2s-1}}w_{2s-1})}{c_s(w_{2s})}
\nonumber
\\
&\times&
\prod_{s=1}^{j-1}(1-q/w_{2s}w_{2s+1})
\prod_{s=1}^j (1-qw_{2s-1}w_{2s})
\prod_{s=1}^{j}H_{\epsilon_{2s-1}}^{2s-1}(w_{2s-1},w_{2s})
\prod_{s=1}^{j-1}H^{2s}(w_{2s},w_{2s+1})\nonumber\\
&\times& J^{2j}(w_{2j})
\frac{1}{D(z,w_1)}=0.
\label{induction:16}
\end{eqnarray}
Using
(\ref{eqn:recursion2}), (\ref{eqn:recursion2}),  (\ref{eqn:recursion3}), (\ref{eqn:recursion4}), (\ref{eqn:recursion5}), 
(\ref{eqn:recursion6}), (\ref{eqn:recursion7}),
the relation (\ref{induction:16}) is shown in the same way as Proposition \ref{prop:proof1}.
~~~{\bf Q.E.D.}

\section{Concluding remark}
\label{sec:7}

In the present paper, 
the $U_q(\widehat{gl}(N|N))$-analog 
of the half-infinite $t$-$J$ model with a diagonal boundary
is considered by using the vertex operator approach.
A bosonization of the boundary state ${_B\langle i|}$ satisfying
${_B\langle i|}T_B(z)={_B \langle i|}$
is constructed
by acting exponential of the bosonic operator $G$ on the highest-weight vector $\langle \Lambda_i|$ 
in the integrable highest-weight module $V^*(\Lambda_i)$ :
\begin{eqnarray}
{_B\langle i|}=\langle \Lambda_i |e^{G} \cdot {Pr},\nonumber
\end{eqnarray} 
where $Pr$ is the projection operator.
In the present paper we focus our attention to $V^*(\Lambda_i)$ for $i=0$ and $i=2N-1$.
The boundary states in $V^*(\Lambda_i)$ for $i=1,2$ can be constructed in the same way.

For more general integrable boundary conditions,
bosonizations of the boundary states are open problem.
Here we study
non-diagonal solutions of the boundary Yang-Baxter equation associated with the quantum superalgebra $U_q(\widehat{gl}(N|N))$.
Let us set $V^{(+)}=\oplus_{j=1}^{N}{\bf C}v_{2j-1}$ and $V^{(-)}=\oplus_{j=1}^N {\bf C}v_{2j}$.
We have $V=V^{(+)} \oplus V^{(-)}$.
Let us study the boundary Yang-Baxter equation in $V \otimes V$ with the $R$-matrix (\ref{def:R-matrix}).
\begin{eqnarray}
\overline{K}_2(z_2)\overline{R}_{2,1}(z_1z_2)
\overline{K}_1(z_1)\overline{R}_{1,2}(z_1/z_2)=
\overline{R}_{2,1}(z_1/z_2)\overline{K}_1(z_1)
\overline{R}_{1,2}(z_1z_2)\overline{K}_2(z_2).
\nonumber
\end{eqnarray}
Let us set $\overline{K}(z)v_i=\sum_{j=1}^{2N}v_j \overline{K}_j^i(z)$.
For $a \neq b$ we have
\begin{eqnarray}
&&
\left(\overline{K}_2(z_2)\overline{R}_{2,1}(z_1z_2)
\overline{K}_1(z_1)\overline{R}_{1,2}(z_1/z_2)\right)_{a,a}^{b,b}=
(-1)^{([v_a]+[v_b])[v_a]}K_a^b(z_2)R_{b,a}^{b,a}(z_1z_2)K_a^b(z_1)R_{b,b}^{b,b}(z_1/z_2),\nonumber\\
&&
\left(\overline{R}_{2,1}(z_1/z_2)\overline{K}_1(z_1)
\overline{R}_{1,2}(z_1z_2)\overline{K}_2(z_2)\right)_{a,a}^{b,b}
=(-1)^{([v_a]+[v_b])[v_a]}R_{a,a}^{a,a}(z_1/z_2)K_a^b(z_1)
R_{b,a}^{b,a}(z_1z_2)K_a^b(z_2).
\nonumber
\end{eqnarray}
Hence we have the following necessary and sufficient condition upon the assumption 
$\overline{K}_a^b(z)\neq 0$.
\begin{eqnarray}
({\rm LHS})_{a,a}^{b,b}=({\rm RHS})_{a,a}^{b,b} \Longleftrightarrow
\overline{R}_{a,a}^{a,a}(z)=\overline{R}_{b,b}^{b,b}(z).
\end{eqnarray}
Because
$\overline{R}_{a,a}^{a,a}(z)\neq
\overline{R}_{b,b}^{b,b}(z)$ for $a \neq b$ $(mod.2)$,
we have
\begin{eqnarray}
a \neq b~(mod.2) \Longrightarrow \overline{K}_a^b(z)=0.
\end{eqnarray}
Hence the boundary Yang-Baxter equation in $V \otimes V$ associated with $U_q(\widehat{gl}(N|N))$ splits into 
two boundary Yang-Baxter equation in $V^{(\pm)} \otimes V^{(\pm)}$ associated with $U_q(\widehat{sl}(N))$.
Hence we get the following procedure to construct non-diagonal solution of the
boundary Yang-Baxter equation associated with $U_q(\widehat{gl}(N|N))$. 
{\it (i) First, we give a diagonal solution in ${\rm End}(V)$ associated with $U_q(\widehat{gl}(N|N))$.
(ii) Next, we extend it to non-diagonal by using two boundary Yang-Baxter equation in $V^{(\pm)} \otimes V^{(\pm)}$ 
associated with $U_q(\widehat{sl}(N))$.}
The same argument holds for $U_q(\widehat{sl}(M|N))$~$(M \neq N)$.

We study
the $U_q(\widehat{gl}(N|N))$-analog 
of the half-infinite $t$-$J$ model with a non-diagonal boundary.
\\
$\bullet$~$U_q(\widehat{gl}(1|1))$ : From the argument above, 
there isn't non-diagonal solution of the boundary Yang-Baxter equation.
There exists diagonal solution only.\\
$\bullet$~$U_q(\widehat{gl}(2|2))$ : 
Let us set $D(z) \in {\rm End}(V)$ be a diagonal solution.
Let us set $O^{(\pm)}(z)$ in ${\rm End}(V^{(\pm)})$ be two off-diagonal solutions associated with $U_q(\widehat{sl}(2))$.
The following $K(z)$ gives a non-diagonal solution associated with $U_q(\widehat{gl}(2|2))$.
\begin{eqnarray}
K(z)=D(z)+O^{(+)}(z)+O^{(-)}(z).
\end{eqnarray}
Here $O^{(\pm)}(z) \in {\rm End}(V^{(\pm)})$ are understood as operators in ${\rm End}(V)$.
For {\it triangular boundary condition},
we have progress on the boundary state associated with $U_q(\widehat{sl}(2))$ \cite{BB, BK1, BK2}.
We would like to report the boundary state of $U_q(\widehat{gl}(2|2))$ spin chain with a triangular boundary
in another paper.

\subsection*{Acknowledgements}
This work is supported by the Grant-in-Aid for 
Scientific Research {\bf C} (26400105)
from Japan Society for Promotion of Science.
The author would like to thank Professor Pascal Baseilhac for discussions.


\begin{appendix}

\section{Quantum superalgebra $U_q(\widehat{gl}(N|N))$}
\label{app:1}

In this Appendix we give the definition of the quantum superalgebra $U_q(\widehat{gl}(N|N))$ \cite{Z}.
We introduce the enlarged Cartan matrix $A=(A_{i,j})_{0 \leq i,j \leq 2N}$ for $\widehat{gl}(N|N)$ as follows.
Let $\{\alpha_i|i=0,1,2,\cdots,2N-1\}$ a set of simple roots of the quantum superalgebra
$\widehat{sl}(N|N)$ given by
\begin{eqnarray}
\alpha_0=\delta-\epsilon_1+\epsilon_{2N},~~~
\alpha_j=\epsilon_j-\epsilon_{j+1}~~~(j=1,2,\cdots,2N-1),
\end{eqnarray}
where we have set $\{\epsilon_j\}_{1\leq j \leq 2N}$, $\delta$ satisfying
$(\delta|\delta)=0, (\delta|e_j)=0$, and $(\epsilon_i|\epsilon_j)=(-1)^{j+1}\delta_{i,j}$.
The enlarged Cartan matrix $(A_{i,j})_{0\leq i,j \leq 2N-1}$ for $\widehat{sl}(N|N)$
is given by $A_{i,j}=(\alpha_i|\alpha_j)$ $(0\leq i,j\leq 2N-1)$.
We extend $\widehat{sl}(N|N)$ by adding the element $\alpha_{2N}=\sum_{j=1}^{2N}\epsilon_j$.
The enlarged Cartan matrix $A$ for
$\widehat{gl}(N|N)$ is given
by $A_{i,j}=(\alpha_i|\alpha_j)$ $(0\leq i,j\leq 2N)$.
For $N=1$ we have
\begin{eqnarray}
A=(A)_{0 \leq i,j \leq 2}=\left(
\begin{array}{ccc}
0&0&-2\\
0&0&2\\
-2&2&0
\end{array}\right).
\label{def:cartan1}
\end{eqnarray}
For $N=2,3,4,\cdots$, we have
\begin{eqnarray}
A=(A_{i,j})_{0\leq i,j \leq 2N}=\left(\begin{array}{ccccccc}
0&-1&0&\cdots &0&1&-2\\
-1&0&1&0&\cdots &0&2\\
0&1&0&-1&\cdots&\cdots &-2\\
\cdots&\cdots&\cdots&\cdots&\cdots&\cdots&\cdots\\
0&\cdots&\cdots&1&0&-1&-2\\
1&0&\cdots&0&-1&0&2\\
-2&2&-2&\cdots&-2&2&0
\end{array}\right).
\label{def:cartan2}
\end{eqnarray}
Note that the Cartan matrix $\bar{A}=(A_{i,j})_{1\leq i,j \leq 2N}$ is invertible.

\begin{dfn}~~~The quantum superalgebra $U_q(\widehat{gl}(N|N))$ $(N=1,2,3,\cdots)$ is generated by the Chevalley generators
$\{e_i,f_i,h_j,d|i=0,1,\cdots,2N-1, j=0,1,2,\cdots,2N\}$.
The ${\bf Z}_2$-grading of the Chevalley generators is
$[e_i]=[f_i]=1$ $(i=0,1,2,\cdots,2N-1)$ and zero otherwise.
The defining relations are
\begin{eqnarray}
&&[h_i,h_j]=0~~~(0\leq i,j \leq 2N),\nonumber
\\
&&[h_i,e_j]=A_{i,j}e_j,~~~[d,e_j]=\delta_{j,0}e_j~~~(0\leq i \leq 2N, 0\leq j \leq 2N-1),
\nonumber\\
&&[h_i,f_j]=-A_{i,j}f_j,~~~[d,f_j]=-\delta_{j,0}f_j~~~(0\leq i \leq 2N, 0\leq j \leq 2N-1),
\nonumber\\
&&[e_i,f_j]=\delta_{i,j}\frac{q^{h_i}-q^{-h_i}}{q-q^{-1}}~~~(1\leq i,j \leq 2N-1),
\nonumber\\
&&[e_i,e_j]=[f_i,f_j]=0~~~{\rm for}~A_{i,j}=0,
\nonumber\\
&&[[e_0,e_1]_{q^{-1}},[e_0,e_{2N-1}]_q]=0,
\nonumber\\
&&[[e_j,e_{j-1}]_{q^{(-1)^j}},[e_j,e_{j+1}]_{q^{(-1)^{j+1}}}]=0~~~(1\leq j \leq 2N-2),
\nonumber\\
&&[[e_{2N-1},e_{2N-2}]_{q^{-1}},[e_{2N-1},e_0]_q]=0,
\nonumber\\
&&[[f_0,f_1]_{q^{-1}},[f_0,f_{2N-1}]_q]=0,
\nonumber\\
&&[[f_j,f_{j-1}]_{q^{(-1)^j}},[f_j,f_{j+1}]_{q^{(-1)^{j+1}}}]=0~~~(1\leq j \leq 2N-2),
\nonumber
\\
&&[[f_{2N-1},f_{2N-2}]_{q^{-1}},[f_{2N-1},f_0]_q]=0.
\end{eqnarray}
Here and throughout this paper, we use 
\begin{eqnarray}
[a,b]_x=ab-(-1)^{[a][b]}x ba.
\end{eqnarray}
The multiplication rule on the tensor products  is ${\bf Z}_2$-graded.
\begin{eqnarray}
(a \otimes b) (a'\otimes b')=(-1)^{[b][a']}(a a' \otimes b b').
\end{eqnarray}
The quantum superalgebra $U_q(\widehat{gl}(N|N))$ has the ${\bf Z}_2$-graded Hopf algebra structure
with the following coproduct $\Delta$, counit $\epsilon$, and antipode $S$.
\begin{eqnarray}
&&\Delta(h_i)=h_i\otimes 1+1\otimes h_i,\nonumber
\\
&&\Delta(e_j)=e_j \otimes 1+q^{h_j}\otimes e_j,~~~
\Delta(f_j)=f_j \otimes q^{-h_j}+1\otimes f_j,\nonumber
\\ 
&&
\epsilon(e_j)=\epsilon(f_j)=\epsilon(h_i)=0,\nonumber
\\
&&S(e_i)=-q^{-h_i}e_i,~~~
S(f_j)=-f_j q^{h_j},~~~S(h_i)=-h_i.
\end{eqnarray}
where $i=0,1,2,\cdots,2N$ and $j=0,1,2,\cdots, 2N-1$.
The coproduct $\Delta$ satisfies algebra automorphism $\Delta(ab)=\Delta(a)\Delta(b)$ and
the antipode satisfies ${\bf Z}_2$-graded algebra anti-automorphism 
$S(ab)=(-1)^{[a][b]}S(b)S(a)$.
\end{dfn}

We denote by ${\cal H}=\oplus_{j=0}^{2N}{\bf C}h_j \oplus {\bf C}d$ 
the extended Cartan subalgebra.
Let $\{\Lambda_0,\Lambda_1,\cdots,\Lambda_{2N},\delta\}$ be the dual basis with $\Lambda_i$ being fundamental weight. Explicitly
\begin{eqnarray}
(\Lambda_i|h_j)=\delta_{i,j},~~~(\Lambda_i|\delta)=0,~~~(d|h_j)=0,~~~(\delta|d)=1~~~(0\leq i,j \leq 2N).
\end{eqnarray}

The quantum superalgebra $U_q(\widehat{gl}(N|N))$ has another realization that we call the Drinfeld realization.

\begin{thm}~~~
The quantum superalgebra $U_q(\widehat{gl}(N|N))$ $(N=1,2,3,\cdots)$ is generated by the Drinfeld generators
$\{X_m^{\pm, i}, H_m^j, c, d|m \in {\bf Z}, i=1,2, \cdots,2N-1, j=1,2,\cdots,2N\}$.
The ${\bf Z}_2$-grading of the Drinfeld generators is
$[X_m^{\pm, i}]=1$ $(i=1,2,\cdots,2N-1)$ and zero otherwise.
The defining relations are
\begin{eqnarray}
&&[c,a]=[d,H_0^j]=[H_0^i,H_n^j]=0~~~(a \in U_q(\widehat{gl}(N|N)), 1 \leq i,j \leq 2N, n \in {\bf Z}),
\nonumber\\
&&[H_m^i,H_n^j]=\delta_{m+n,0}\frac{[A_{i,j}m]_q [cm]_q}{m}~~~~(m,n \in {\bf Z}_{\neq 0}, 1\leq i,j \leq 2N),
\nonumber\\
&&[H_0^i, X_j^\pm(z)]=\pm A_{i,j} X_j^\pm(z)~~~(1\leq i \leq 2N, 1\leq j \leq 2N-1),
\nonumber\\
&&[H_m^i, X_j^\pm(z)]=\pm \frac{[A_{i,j}m]_q}{m}q^{\mp c|m|}z^m X_j^\pm(z)~~~(1\leq i \leq 2N, 1\leq j \leq 2N-1, m \in {\bf Z}_{\neq 0}),
\nonumber\\
&&[X_i^+(z_1),X_j^-(z_2)]\nonumber\\
&=&
\frac{\delta_{i,j}}{(q-q^{-1})z_1z_2}
\left(\delta(q^{-c}z_1/z_2)\psi_i^+(q^{\frac{c}{2}}z_2)-\delta(q^cz_1/z_2)\psi_i^-(q^{-\frac{c}{2}}z_2)\right)~~~
(1\leq i,j \leq 2N-1),
\nonumber\\
&&[X_i^\pm(z_1),X_j^\pm(z_2)]=0~~~(A_{i,j}=0),
\nonumber\\
&&
(z_1-q^{\pm A_{i,j}}z_2)X_i^\pm(z_1)X_j^\pm(z_2)=(q^{\pm A_{j,i}}z_1-z_2)X_j^\pm(z_2)X_i^\pm(z_1)~~~(A_{i,j}\neq 0),
\nonumber\\
&&
\left[[X_j^\pm(z_1),X_{j-1}^\pm(z_2)]_{q^{(-1)^j}},
[X_j^\pm(z_3),X_{j+1}^\pm(z_4)]_{q^{(-1)^{j+1}}}\right]\nonumber\\
&&+\left[[X_j^\pm(z_3),X_{j-1}^\pm(z_2)]_{q^{(-1)^j}},
[X_j^\pm(z_1),X_{j+1}^\pm(z_4)]_{q^{(-1)^{j+1}}}\right]=0~~~(2 \leq j \leq 2N-2).
\end{eqnarray}
Here we have used the generating functions
\begin{eqnarray}
&&
X_j^\pm(z)=\sum_{m \in {\bf Z}}X_{m}^{\pm, j}z^{-m-1}~~~(1\leq j \leq 2N-1),
\nonumber\\
&&
\psi_j^\pm(z)=q^{\pm H_0^j}\exp\left(\pm (q-q^{-1})\sum_{m=1}^\infty
H_m^j z^{\mp m}\right)~~~(1 \leq j \leq 2N-1).
\end{eqnarray}
The Chevalley generators are related to the Drinfeld generators as follows.
\begin{eqnarray}
&&
h_0=c-\sum_{j=1}^{2N-1}H_0^j,~~~
h_i=H_0^i,~~e_j=X_0^{+,j},~~~f_j=X_0^{-,j}~~~(1\leq i \leq 2N, 1\leq j \leq 2N-1),\nonumber\\
&&
e_0=[X_0^{-,2N-1},[X_0^{-,2N-2},\cdots,[X_0^{-,3},[X_0^{-,2},X_1^{-,1}]_q]_{q^{-1}}\cdots ]_q]_{q^{-1}} q^{h_0-c},
\nonumber\\
&&
f_0=(-1)^N q^{c-h_0}[[\cdots[[X_{-1}^{+,1},X_0^{+,2}]_{q^{-1}},X_0^{+,3}]_q,\cdots,X_0^{+,2N-2}]_{q^{-1}},X_0^{+,2N-1}]_q.
\end{eqnarray}
\end{thm}

\section{Boundary state in $V(\Lambda_{2N-1})$}
\label{app:2}

In this Appendix we give a bosonization of the boundary state in the integrable highest-weight module $V(\Lambda_{2N-1})$.
Let $L>0, M, R \geq 0$ $(L+M+R=2N)$.
The boundary $K$-matrix $K(z)$ is given by
$K(z)=\frac{\varphi(z)}{\varphi(z^{-1})} \overline{K}(z)$.
Here the matrix $\overline{K}(z)$ is given by
\begin{eqnarray}
\overline{K}(z)=
{\rm diag}\left(\underbrace{1,\cdots,1}_{L},
\underbrace{\frac{1-r/z}{1-rz},\cdots,\frac{1-r/z}{1-rz}}_{M},
\underbrace{z^{-2},\cdots,z^{-2}}_{R}\right).
\end{eqnarray}
Here the function $\varphi(z)$ is given by
\begin{eqnarray}
\varphi(z)=\left\{\begin{array}{cc}
\varphi^{[1]}(z)& (L>0, M=R=0)\\
\varphi^{[2]}(z)& (L, M>0,R=0)\\
\varphi^{[3]}(z)& (L, M, R>0)
\end{array}
\right.,
\end{eqnarray}
where we have set
\begin{eqnarray}
&&\varphi^{[1]}(z)=(1-qz^2)^{\frac{1-N}{2N}}(1+z^2)^{\frac{1-2N}{2N}},
\\
&&\varphi^{[2]}(z)=\varphi^{[1]}(z)\times\left\{\begin{array}{cc}
(1-rz/q)^{\frac{1-2N}{2N}}& (L={\rm odd})\\
1&(L={\rm even})\end{array}\right.,
\\
&&\varphi^{[3]}(z)=
\varphi^{[2]}(z) 
\times\left\{
\begin{array}{cc}
(1-qz/r)^{\frac{1-2N}{2N}} & (L={\rm odd}, M={\rm even})\\
(1-z/rq)^{\frac{1-2N}{2N}}& (L={\rm even}, M={\rm odd})\\
1 & ({\rm otherwise})
\end{array}\right..
\end{eqnarray}

\begin{thm}~~~The boundary state ${_B \langle 2N-1|} \in V^*(\Lambda_{2N-1})$ is realized as follows.
\begin{eqnarray}
{_B \langle 2N-1|}=\langle \Lambda_{2N-1}|e^G \cdot Pr,
\end{eqnarray}
where the highest-weight vector $\langle \Lambda_{2N-1}|$ is given in (\ref{lambda2N-1}).
Here the projection operator $Pr$ is given by
$\displaystyle Pr=\prod_{j=1}^{N-1} \eta_0^j \prod_{j=1}^{N-1} \xi_0^j
\cdot \xi_0^{N} \eta_0^N$.
Here the bosonic operator $G$ is given in (\ref{def:G}),
where $\delta_m^j$ and $\beta_m^i$ are given as follows.
\begin{eqnarray}
\delta_m^j&=&
\left\{
\begin{array}{cc}\frac{\displaystyle q^{-m}}{\displaystyle
[m]_q}(1-2(-1)^{\frac{m}{2}})\theta_m & (1\leq j \leq N-1)\\
\frac{\displaystyle q^{-m}}{\displaystyle [m]_q}\theta_m & (j=N)
\end{array}
\right.,
\label{lambda2N-1:delta}
\\
\beta_m^j&=&\left\{\begin{array}{cc}
\beta_m^{[1], j}& (L>0, M=R=0)\\
\beta_m^{[2], j}& (L, M>0,R=0)\\
\beta_m^{[3], j}& (L,M,R>0)
\end{array}\right.~~(i=1,2,\cdots,2N),
\label{lambda2N-1:beta}
\end{eqnarray}
where we have set
\begin{eqnarray}
&&
\beta_m^{[1],2s-1}=\frac{2(1-s)}{[m]_q}q^{-\frac{3}{2}m}\theta_m~~~(1\leq s \leq N),\\
&&
\beta_m^{[1],2s}=
\frac{2(1-s)}{[m]_q}q^{-\frac{3}{2}m}\theta_m-\left\{
\begin{array}{cc}
0& (1\leq s \leq N-1)\\
\frac{2(-1)^{\frac{m}{2}}}{[m]_q}q^{-\frac{3}{2}m}\theta_m
& (s=N)
\end{array}
\right.,
\\
&&
\beta_m^{[2],i}=\beta_m^{[1],i}+\left\{\begin{array}{cc}
0 & (1\leq j \leq M)\\
\frac{\displaystyle r^m q^{-\frac{5}{2}m}}{\displaystyle [m]_q}& (M<i \leq 2N, M={\rm odd})\\
\frac{\displaystyle r^m q^{-\frac{3}{2}m}}{\displaystyle [m]_q}& (M<i \leq 2N, M={\rm even})
\end{array}\right.,
\\
&&
\beta_m^{[3],i}=
\beta_m^{[2],i}+
\left\{\begin{array}{cc}
0& (1\leq i \leq L+M)\\
\frac{\displaystyle r^{-m}q^{\frac{1}{2}m}}{\displaystyle [m]_q}& (L+M<i \leq 2N, L={\rm odd}, M={\rm odd})\\
\frac{\displaystyle r^{-m}q^{-\frac{1}{2}m}}{\displaystyle [m]_q}& (L+M<i \leq 2N, L={\rm odd}, M={\rm even})\\
\frac{\displaystyle r^{-m}q^{-\frac{5}{2}m}}{\displaystyle [m]_q}& (L+M<i \leq 2N, L={\rm even}, M={\rm odd})\\
\frac{\displaystyle r^{-m}q^{-\frac{3}{2}m}}{\displaystyle [m]_q}& (L+M<i \leq 2N, L={\rm even}, M={\rm even})
\end{array}\right..
\end{eqnarray}
\end{thm}

\begin{prop}~~~
We have
\begin{eqnarray}
c_j(w)=\left\{\begin{array}{cc}
1+w^2 &(1\leq j \leq N-1)\\
1&(j=N)
\end{array}\right.,~~~
f_1(z)=\left\{\begin{array}{cc}
\varphi^{[1]}(z)& (L>0, M=R=0)\\
\varphi^{[2]}(z)& (L,M>0, R=0)\\
\varphi^{[3]}(z)& (L,M,R>0)
\end{array}
\right.,
\end{eqnarray}
and
\begin{eqnarray}
g_i(w)&=&\left\{\begin{array}{cc}
g_i^{[1]}(w)& (L>0, M=R=0)\\
g_i^{[2]}(w)& (L,M>0,R=0)\\
g_i^{[3]}(w)& (L,M,R>0)
\end{array}
\right.~~~(i=1,2,\cdots,2N-1),
\end{eqnarray}
where we have set
\begin{eqnarray}
g_i^{[1]}(w)&=&\left\{\begin{array}{cc}
1&(i={\rm odd}, 1\leq i \leq 2N-3)\\
(1-w^2)& (i={\rm even}, 1\leq i \leq 2N-2)\\
(1+w^2)& (i=2N-1)
\end{array}\right.,\\
g_i^{[2]}(w)&=&g_i^{[1]}(w) \times \left\{\begin{array}{cc}
\frac{1}{(1-rw/q)}& (i=L, L={\rm odd})\\
\frac{1}{(1-rw)}& (i=L, L={\rm even})\\
1 & ({\rm otherwise})
\end{array}
\right.,\\
g_i^{[3]}(w)&=&g_i^{[2]}(w)
\times
\left\{
\begin{array}{cc}
\frac{1}{(1-q^2 w/r)}&~~~(i=L+M, L={\rm odd}, M={\rm odd})
\\
\frac{1}{(1-q w/r)}&~~~(i=L+M, L={\rm odd}, M={\rm even})
\\
\frac{1}{(1-w/rq)}&~~~(i=L+M, L={\rm even}, M={\rm odd})
\\
\frac{1}{(1-w/r)}&~~~(i=L+M, L={\rm even}, M={\rm even})
\\
1 & ({\rm otherwise})
\end{array}
\right..
\end{eqnarray}
\end{prop}

\section{Normal ordering}
\label{app:3}

In this Appendix we summarize the normal orderings.
\begin{eqnarray}
\phi_1^*(z)X_{\epsilon}^{-,1}(qw)&=&
:\phi_1^*(z)X_{\epsilon}^{-,1}(qw)
:\frac{-1}{qz(1-qw/z)},
\nonumber\\
X_{\epsilon}^{-,1}(qw)\Phi_1^*(z)&=&
:
X_{\epsilon}^{-,1}(qw)\Phi_1^*(z):\frac{1}{qw(1-qz/w)},
\nonumber
\\
X_\epsilon^{-,2j-1}(qw_1)X^{-,2j}(qw_2)&=&
:X_\epsilon^{-,2j-1}(qw_1)X^{-,2j}(qw_2):
q^{-\epsilon}\frac{(1-qw_2/w_1)}{(1-w_2/q^{\epsilon}w_1)}~~~(1\leq j \leq N-1),\nonumber
\nonumber
\\
X^{-,2j}(qw_2)X_\epsilon^{-,2j-1}(qw_1)&=&
-:X^{-,2j}(qw_2)X_\epsilon^{-,2j-1}(qw_1)
:\frac{(1-qw_1/w_2)}{(1-q^\epsilon w_1/w_2)}~~~(1\leq j \leq N-1),
\nonumber\\
X^{-,2j}(qw_1)X_\epsilon^{-,2j+1}(qw_2)&=&:
X^{-,2j}(qw_1)X_\epsilon^{-,2j+1}(qw_2)
:\frac{1}{qw_1(1-qw_2/w_1)}~~~(1\leq j \leq N-1),\nonumber
\\
X_\epsilon^{-,2j+1}(qw_2)X^{-,2j}(qw_1)&=&:
X_\epsilon^{-,2j+1}(qw_2)X^{-,2j}(qw_1)
:\frac{1}{qw_2(1-qw_1/w_2)}~~~(1\leq j \leq N-1),
\nonumber\\
X^{-,2j-1}(qw_1)X^{-,2j+1}(qw_2)&=&:
X^{-,2j-1}(qw_1)X^{-,2j+1}(qw_2)
:~~~(1\leq j \leq N-1),
\nonumber\\
X^{-,2j+1}(qw_2)X^{-,2j-1}(qw_1)&=&
-:
X^{-,2j+1}(qw_2)X^{-,2j-1}(qw_1):~~~(1\leq j \leq N-1).
\end{eqnarray}
\begin{eqnarray}
e^{B_+(z;\frac{1}{2})}e^{-H_-^1(w;\frac{1}{2})}&=&
\frac{1}{(1-qw/z)}
:e^{-H_-^1(w;\frac{1}{2})}
e^{B_+(z;\frac{1}{2})}:,\nonumber
\\
e^{-H_+^{2j-1}(w_1;\frac{1}{2})}e^{-H_-^{2j}(w_2;\frac{1}{2})}
&=&(1-qw_2/w_1)
:e^{-H_-^{2j}(w_2;\frac{1}{2})}e^{-H_+^{2j-1}(w_1;\frac{1}{2})}:~~~(1\leq j \leq N-1),
\nonumber
\\
e^{-H_+^{2j}(w_1;\frac{1}{2})}e^{-H_-^{2j+1}(w_2;\frac{1}{2})}&=&
\frac{1}{(1-qw_2/w_1)}
:e^{-H_-^{2j+1}(w_2;\frac{1}{2})}e^{-H_+^{2j}(w_1;\frac{1}{2})}:~~~(1\leq j \leq N-1),
\nonumber
\\
e^{-c_+^j(q^{1-\epsilon}/w_1)}e^{c_-^j(qw_2)}&=&
\frac{1}{(1-q^\epsilon w_1w_2)}
:e^{c_-^j(qw_2)}e^{-c_+^j(q^{1-\epsilon}/w_1)}:~~~(1\leq j \leq N).
\end{eqnarray}

\end{appendix}

\end{document}